\documentclass[review]{elsarticle}

\usepackage{lineno,hyperref}
\usepackage{amsmath}
\modulolinenumbers[1]

\usepackage{subcaption}
\usepackage{multirow}
\usepackage{xcolor}
\usepackage{ragged2e}
\usepackage{lineno}
\usepackage{siunitx}
\usepackage{amssymb}
\usepackage{ulem}
\makeatletter
\g@addto@macro\endfrontmatter{\enlargethispage{-2\baselineskip}}
\makeatother

\journal{NIM - A}

\begin{document}

\begin{frontmatter}

\title{The Light Source of the TRIDENT Pathfinder Experiment}

\author[tdli]{Wenlian Li}

\author[sjtu_en]{Xiaohui Liu}
\author[tdli]{Wei Tian}
\author[tdli]{Fuyudi Zhang}
\author[tdli]{Shishen Xian}
\author[sjtu]{Mingxin Wang}
\author[sjtu]{Jiannan Tang}
\author[pku]{Fan Hu}
\author[tdli]{Ziping Ye}
\author[ustc]{Peng Miao}
\author[sjtu]{Zhengyang Sun}

\author[tdli,sjtu]{Donglian Xu\corref{corresponding_author}}
\ead{donglianxu@sjtu.edu.cn}

\cortext[corresponding_author]{Corresponding author}
\address[tdli]{Tsung-Dao Lee Institute, Shanghai Jiao Tong University, Shanghai 201210, China}
\address[sjtu]{School of Physics and Astronomy, Shanghai Jiao Tong University, Key Laboratory for Particle Astrophysics and Cosmology (MoE), Shanghai Key Laboratory for Particle Physics and Cosmology, Shanghai 200240, China} 
\address[sjtu_en]{School of Naval Architecture, Ocean and Civil Engineering, Shanghai Jiao Tong University, Shanghai 200240, China}
\address[pku]{Department of Astronomy, School of Physics, Peking University, Beijing 100871, China}
\address[ustc]{Department of Modern Physics, University of Science and Technology of China, Hefei 230026, China}

\begin{abstract}

In September 2021, a site scouting mission known as the TRIDENT pathfinder experiment (TRIDENT EXplorer, T-REX for short) was conducted in the South China Sea with the goal of envisaging a next-generation multi-cubic-kilometer neutrino telescope. One of the main tasks is to measure the \textit{in-situ} optical properties of seawater at depths between $2800~\mathrm{m}$ and $3500~\mathrm{m}$, where the neutrino telescope will be instrumented. To achieve this, we have developed a light emitter module equipped with a clock synchronization system to serve as the light source, which could be operated in pulsing and steady modes. Two light receiver modules housing both photomultiplier tubes (PMTs) and cameras are employed to detect the photons emitted by the light source. This paper presents the instrumentation of the light source in T-REX, including its design, calibration, and performance.

\end{abstract}


\end{frontmatter}


\section{Introduction}
\label{sec:intro}

Neutrinos serve as exceptional probes of the most extreme environments in the Universe. Being extremely light and participating only in weak and gravitational interactions, neutrinos can escape from dense environments and provide insight into the sources. 
Without being deflected by the interstellar magnetic fields, neutrinos are ideal messengers for unveiling the longstanding enigma of the origins of cosmic rays.
To detect neutrinos, a large volume of detector array consisting of photosensors built in transparent media such as ice and water is required. 
The direction and energy of neutrinos can be reconstructed by collecting the Cherenkov light radiated by secondary charged particles induced from neutrino interactions. 
In 2013, the IceCube Neutrino Observatory made the first detection of an extraterrestrial diffuse neutrino flux \cite{IceCube:2013}. 
Over the past decade, IceCube has observed compelling evidence of neutrino emissions from active galaxies including TXS 0506+056 and NGC 1068 \cite{IceCube:2018, IceCube:2022der}.
To detect astrophysical neutrinos more effectively and confirm their sources, it is necessary to significantly boost the sensitivity of the next-generation neutrino telescopes and grow a neutrino telescope network of global coverage.

\begin{figure}[htbp]
\centering 
\includegraphics[scale=0.43]{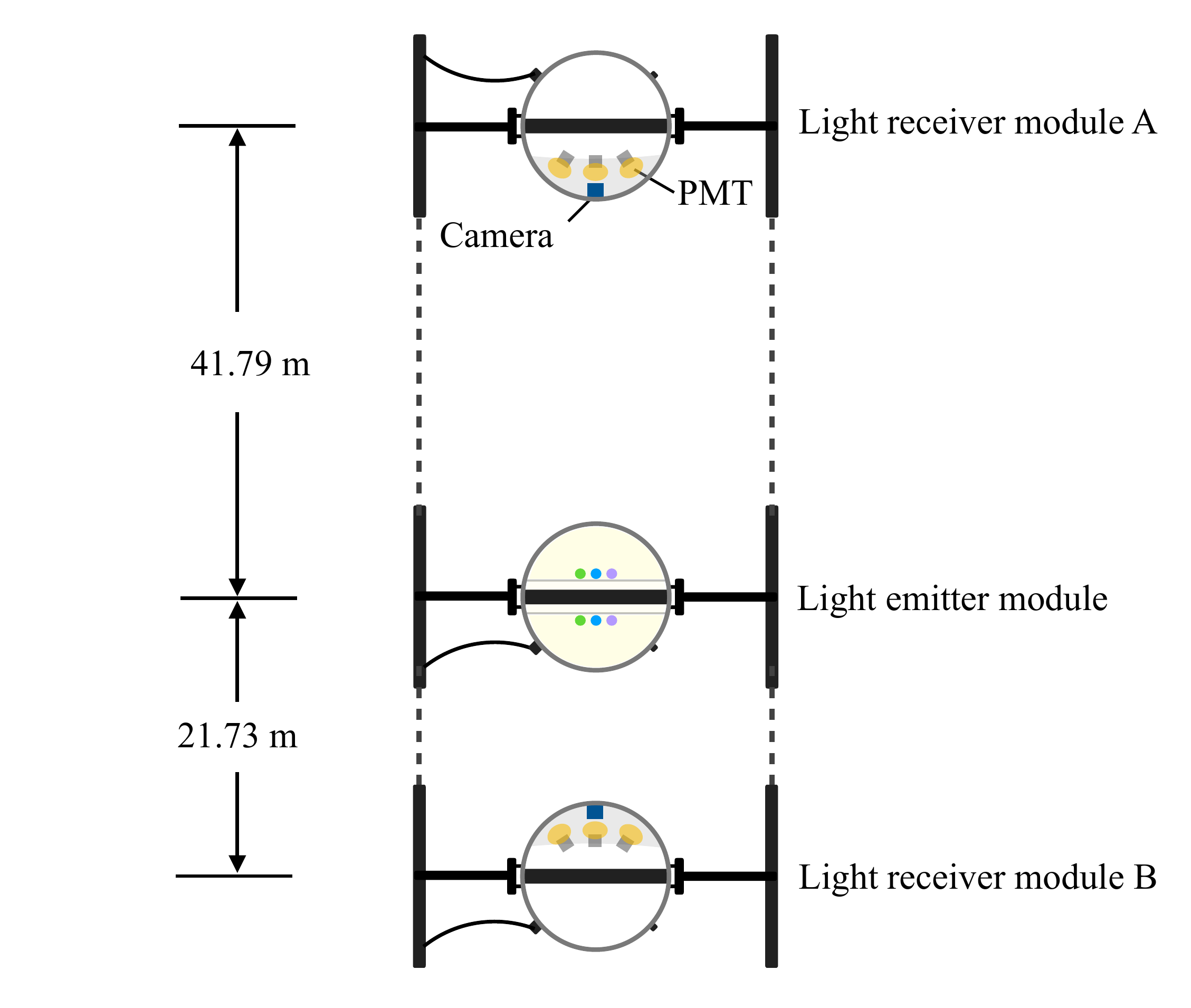}
\caption{\label{fig_TRIDENT_pathfinder_setup} 
A schematic illustration of the geometric arrangement of the three modules in T-REX, comprising a light receiver module A placed at the top, a light receiver module B at the bottom, and a light emitter module located at distances of $41.79\pm 0.04~\mathrm{m}$ and $21.73\pm 0.02~\mathrm{m}$ from modules A and B, respectively. This figure is not drawn to scale.
}
\end{figure}

The tRopIcal DEep-sea Neutrino Telescope (TRIDENT) is a next-generation neutrino telescope to be built in the South China Sea \cite{trident:2022}. As a pathfinder of TRIDENT, T-REX was conducted for the measurements of deep-sea water transmission properties and oceanographic conditions. The T-REX apparatus was deployed to a depth of $3420~\mathrm{m}$, followed by $\sim 2$ hours of operation. It consists of a vertical mooring line instrumenting a light emitter module in the middle and two light receiver modules, A and B, located at the top and bottom (as shown in Figure~\ref{fig_TRIDENT_pathfinder_setup}). The three modules are battery-powered and controlled by the data acquisition system (DAQ) onboard the research vessel.
Two light receiver modules, each housing three 3-inch PMTs and one camera, are used to detect photons emitted from the light source at distances of $21.73\pm 0.02~\mathrm{m}$ and $41.79\pm 0.04~\mathrm{m}$, respectively.
With two independent detection systems of the PMT and camera, the light source has two light-emitting modes: the pulsing mode ($405~\mathrm{nm}$, $450~\mathrm{nm}$, $525~\mathrm{nm}$) for the former and the steady mode ($405~\mathrm{nm}$, $460~\mathrm{nm}$, $525~\mathrm{nm}$) for the latter.
The emission mode and wavelength of the light source are controlled by the slow control system of the DAQ \cite{Wang:2023rvb}.

This paper is structured as follows: Section~\ref{section2_design_concept} presents the experimental requirements of the light source, while Section~\ref{section3_structure} provides a detailed description of its internal structure. The layout, selection, and calibration of the light-emitting diode (LED) systems are discussed in Section~\ref{sec:led system}. In Section~\ref{sec:low temperature calibration}, we describe the low-temperature calibration of the pulsing and steady light source, and Section~\ref{sec: light source performance} discusses the performance of the light source. Finally, a summary is presented in Section~\ref{sec: summary}.

\section{Experimental requirements for the light source}
\label{section2_design_concept}

Photons propagating in water undergo absorption and scattering processes. 
To measure these optical properties, a relative measurement strategy is adopted. 
This approach largely mitigates hidden systematic uncertainties, including light loss caused by changing refractive indexes at the air-glass-seawater interfaces and the influence of deep-sea water conditions, such as temperature, on the detector's response.
To implement this strategy, the light source aims to supply isotropic light to the PMTs and cameras on both sides, which ensures that the measurement is not influenced by the detector's viewing angle.
Isotropic light sources have been previously developed by ANTARES \cite{ANTARES:2005} and P-ONE STRAW-a (namely POCAM) \cite{STRAW:2019} for measuring seawater's optical properties, with the POCAM initially designed for use in the IceCube Upgrade \cite{Henningsen:2020zsj}.
To perform relative measurements, we calibrated the relative light intensity of both hemispheres of the light source (as presented in Section~\ref{sec:low temperature calibration}) and the relative sensitivity of the PMTs in the light receiver modules (as discussed in \cite{Yudi:2022}). The calibrated results serve as important inputs in the data analyses \cite{Yudi:2022, TianWei:2022} and simulation \cite{Hufan:2022}.
Additionally, we have estimated the impact of the mechanical support cables on the relative light output. The blocking effect of the cable is less than 0.1\% and can be neglected.
Since the Cherenkov spectrum is broad, the light source therefore instruments several different wavelength LEDs to achieve a multi-wavelength coverage of the Cherenkov spectrum in the seawater.
For the PMT system, a pulsing light source that emits intense ($O(10^9)$ photons per pulse) light pulses with a nanosecond width is needed to distinguish between direct and scattered light, considering the light attenuation at long detection distance in seawater. The intensity of light pulses should be adjustable to ensure that PMTs can receive single photoelectron (SPE) signals. For the camera system, a homogeneous steady light source is needed since cameras record images.

To simplify and unify the structure, pulsing and steady LEDs are mounted on one LED board but driven by different driver circuits \cite{Jiannan:2022}.
The isotropy of the light source is optimized by employing a double diffuser structure. 
The internal support structure is symmetrical and is optimized to maximize the light-emitting area while accommodating all electronic boards of the light source.

\section{The light source structure}
\label{section3_structure}

\begin{figure}[htbp]
\centering 
\includegraphics[width=.99\textwidth]{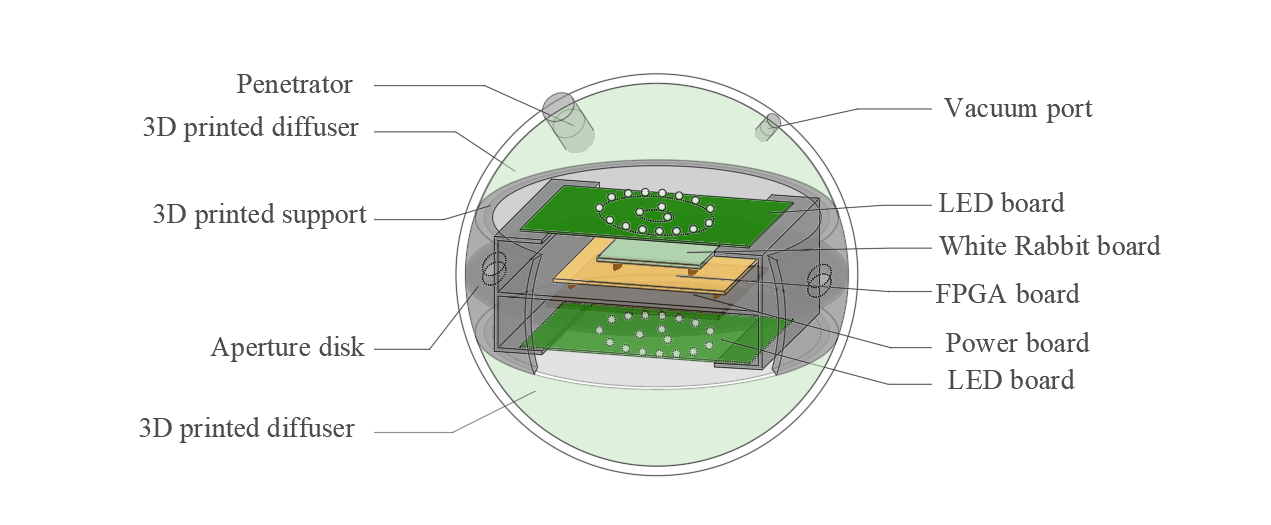}
\caption{\label{fig_housing} 
Sectional view of the light source instrument. The 3D printed support, depicted in middle gray, accommodates all electronic boards. The light green components at the top and bottom denote the 3D printed diffusers.
}
\end{figure}

This section describes the hardware of the light source, including the 3D printed supports, diffusers, and the electronics. These components are integrated into a spherical glass vessel with a diameter of $432~\mathrm{mm}$ and a thickness of $14~\mathrm{mm}$, which serves as the housing for the entire system. A sectional view of the light source is presented in Figure~\ref{fig_housing}, and detailed information on its structure is provided in the following subsections.

\subsection{Instrument housing}

The complete support structure is enclosed within a 17-inch Vitrovex$^\circledR$ glass sphere produced by Nautilus, which can withstand water pressure up to $6700~\mathrm{m}$ deep.
The glass vessel comprises two hemispheres, with one hemisphere featuring two holes for the vacuum port and the penetrator, referred to as the ``hole hemisphere''. The other hemisphere is referred to as the ``complete hemisphere''.
The encapsulation of the entire glass sphere was carried out using the same procedure as that used for the receiver module \cite{Yudi:2022}. 
To verify the performance of the housing under deep-sea conditions, the instrument underwent a series of high-pressure tests conducted at the Underwater Engineering Institute \footnote{https://underwater.sjtu.edu.cn} of Shanghai Jiao Tong University. Multiple tests were performed under water pressures ranging from $38~\mathrm{Mpa}$ to $45~\mathrm{Mpa}$, with durations lasting from $2$ to $24$ hours. In the last few tests before deployment, ice was added to the pressure tank to simulate the deep-sea low-temperature conditions.

\subsection{Diffusers}

\begin{figure}[htbp]
\centering 
\includegraphics[width=.5\textwidth]{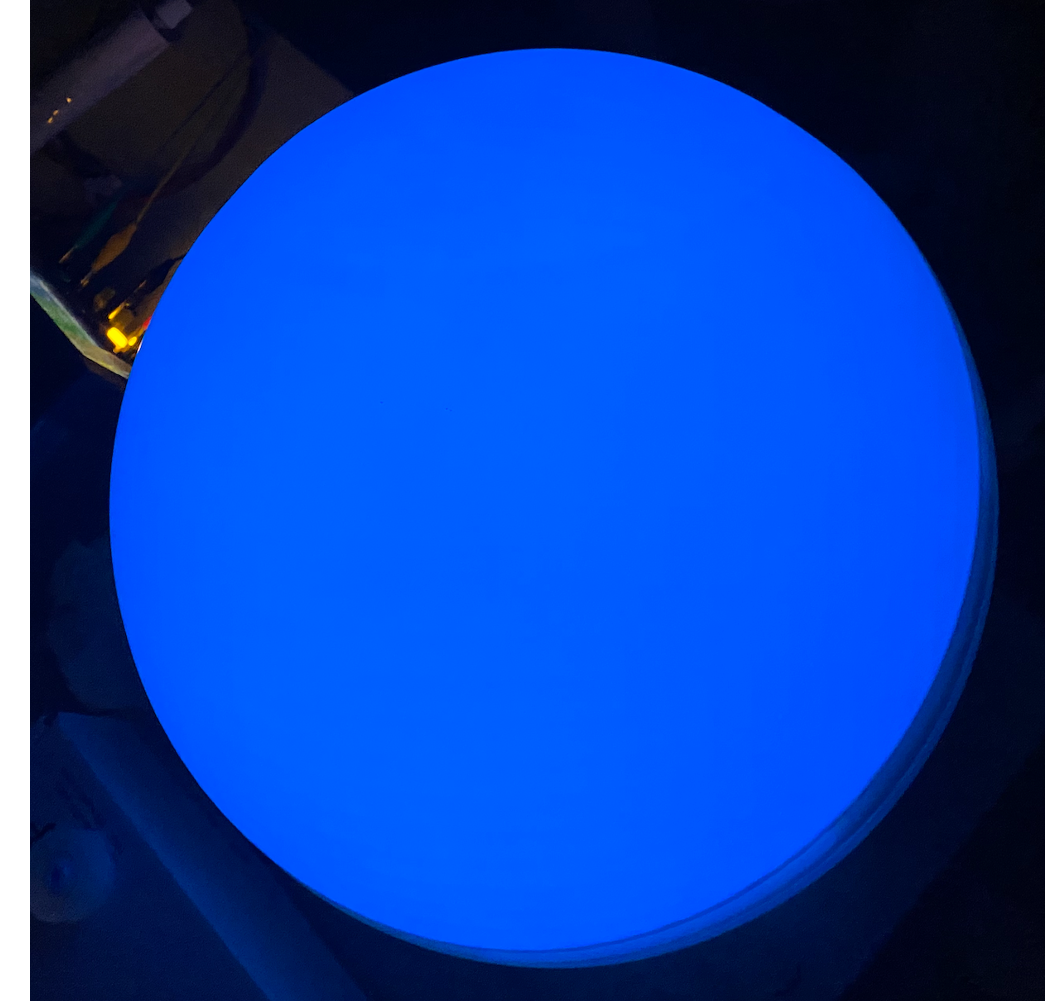}
\caption{\label{fig_double_layer_diffuser}
Image of the 3D printed diffuser with $460~\mathrm{nm}$ LEDs emitting light, which is diffused by Teflon balls and the 3D printed diffuser.
}
\end{figure}

We developed a double diffuser structure to enhance the isotropy of the light source. The first diffuser consists of a Teflon ball made of polytetrafluoroethylene (PTFE). The second diffuser is a 3D-printed spherical shell with a thickness of $1.5~\mathrm{mm}$, which is optimized to achieve a balanced compromise between light transmission and structural stability.
Figure~\ref{fig_double_layer_diffuser} shows the 3D printed diffuser and the light that has been diffused by the double diffuser structure.

\subsection{3D-printed support structure}

The internal support structure of the light source is fabricated using 3D printing technology. This approach offers several advantages, including cost-effectiveness, customization, rapid prototyping, and fast production, making it suitable for producing internal support structures for optical modules. The 3D printing technology has been applied to neutrino telescopes such as KM3NeT \cite{Bruijn:2015huz}.

\begin{figure}[htbp]
\centering 
\includegraphics[width=.98\textwidth]{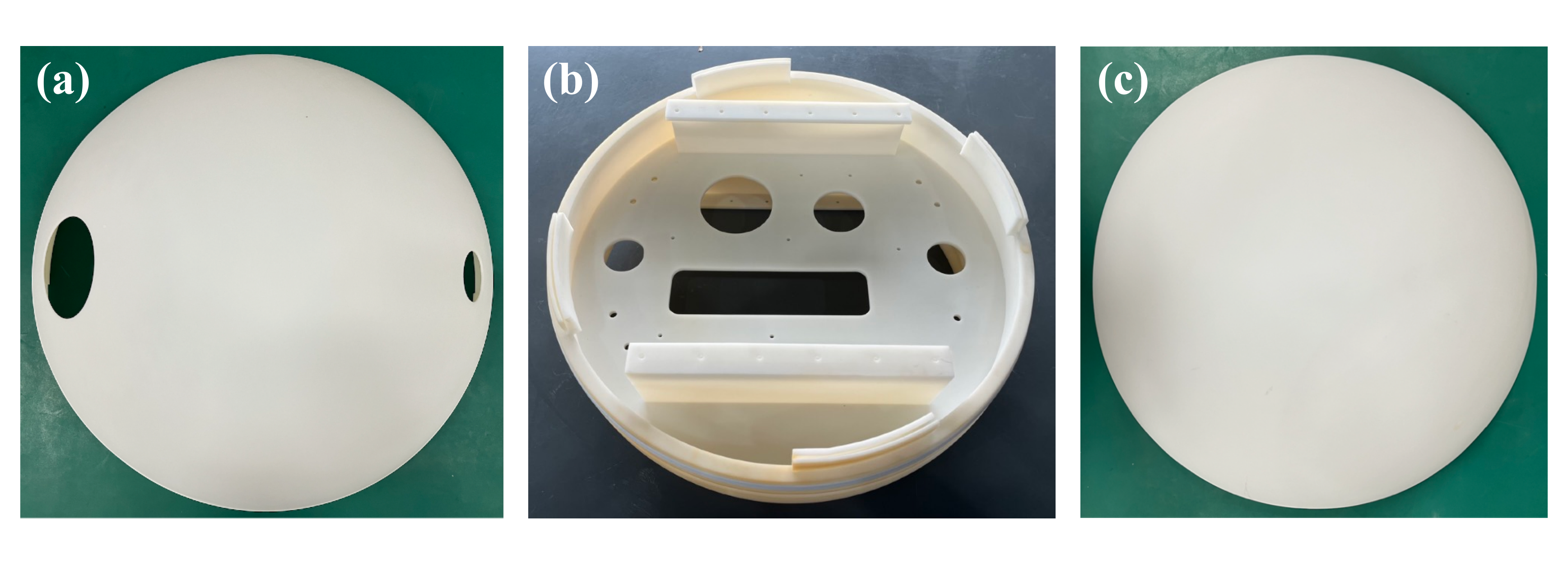}
\caption{\label{fig_steam} 3D printed supports of the light source. Shown are the hole diffuser (a) with two holes for the vacuum port and penetrator, the middle 3D printed support for housing electronic boards (b), and the complete diffuser (c).}
\end{figure}

The 3D printed supports inside the glass sphere comprise three parts: a ring-shaped support in the middle for housing electronic boards and two 3D-printed spherical shells on the top and bottom for diffusing the light emitted by LEDs. 
Figure~\ref{fig_steam} illustrates the 3D printed supports used in T-REX.
The 3D printed diffusers, depicted in Figure~\ref{fig_steam} (a) and (c), are attached to the top and bottom of the middle support shown in Figure~\ref{fig_steam} (b) using a specially designed slotted structure. An air gap exists between the glass vessel and the hemispherical diffusers (a) and (c).
The middle 3D printed support shown in Figure~\ref{fig_steam} (b) features an aperture disk that serves multiple functions. 
It provides screw holes for electronic board installation and allows for the passage of cables, facilitating heat dissipation. Two parallel shelves on both sides of the aperture disk are intended for mounting LED boards, while other electronic boards are placed between them to avoid interference with LED light emission. This support structure includes four annular grooves on the side, which are bonded with foamed silicone strips to provide friction against the inner surface of the glass vessel. A horizontal line is incorporated to align with the equatorial plane of the glass vessel during installation.

The 3D printed supports are made from photosensitive resin, which exhibits excellent moisture resistance and toughness, making it an ideal material for mounting electronic boards. In addition, its good surface smoothness makes it suitable for use as a diffuser. The supports are produced using the mature Stereo lithography Appearance (SLA) technique, which relies on the principle of photo-curing. This technique is capable of producing large-size models with high precision \cite{QUAN2020110}, making it an optimal approach for manufacturing internal supports of optical modules.
It should be noted that the 3D printed supports made from photosensitive resin are sufficient for use in a short-term experiment like T-REX. However, further tests and validations on the stability of the 3D printed materials are needed for long-term applications.

\subsection{Electronics}

The electronics of the light source consist of a power supply system, a slow control system, and a clock synchronization system, as described in~\cite{Wang:2023rvb}. The power board provides controllable power to all electronic boards in the light source, while the central logic board (CLB), which includes a field programmable gate array (FPGA) board and a digital-to-analog mezzanine board (DACB), manages the slow control and media access control (MAC) packets. The White Rabbit (WR) system ensures clock synchronization between the light source and the receiver modules. Two distinct LED driver circuits were developed to correspond with the two light-emitting modes of the light source. Details about the design of the LED drivers are discussed in~\cite{Jiannan:2022}.
For the pulsing LEDs, a ``Kapustinsky's" pulser \cite{1985NIMPA.241..612K} is used to drive nanosecond-width light pulses. The light intensity of the pulsing LEDs is tunable, with an adjustable supply voltage ranging from $0~\mathrm{V}$ to $30~\mathrm{V}$. A step-up DC/DC converter simultaneously drives five identical steady LEDs with a constant current of $20~\mathrm{mA}$, ensuring their brightness is nearly uniform.

\section{The LED system}
\label{sec:led system} 

The light source is equipped with two LED systems that enable two different modes: pulsing and steady. The pulsing mode is achieved using three LEDs with different wavelengths, while the steady mode uses fifteen LEDs, with five identical LEDs for each wavelength.

\subsection{LED layout}
\label{sec:led_layout}

\begin{figure}[htbp]
\captionsetup[subfigure]{justification=Centering}
\begin{subfigure}[t]{0.6\textwidth}
    \includegraphics[width=\linewidth]{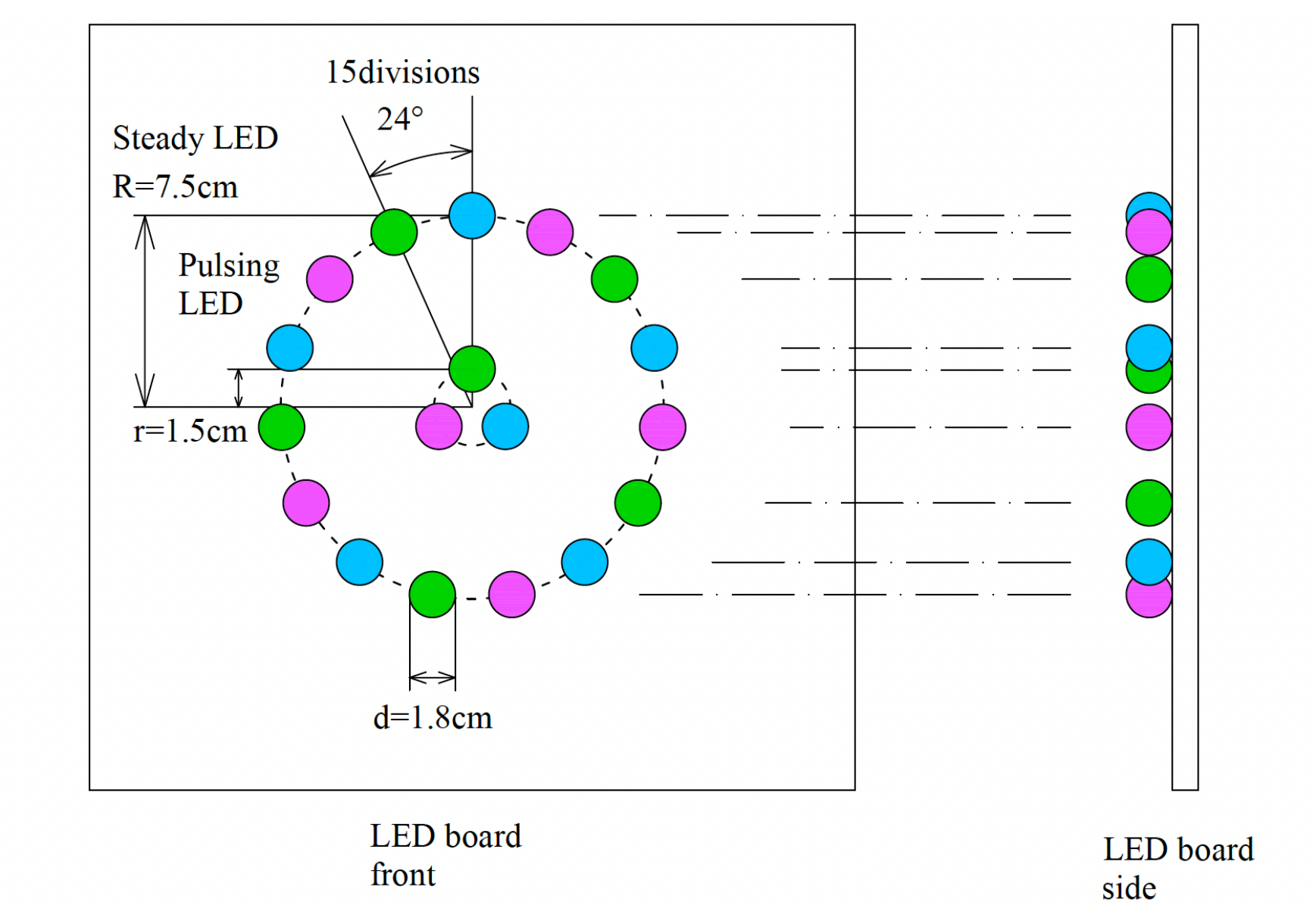}
    \caption{\label{fig_LED_layout_a}Sketch of LED layout.}
\end{subfigure}\hspace{\fill}
\quad
\begin{subfigure}[t]{0.36\textwidth}
    \includegraphics[width=\linewidth]{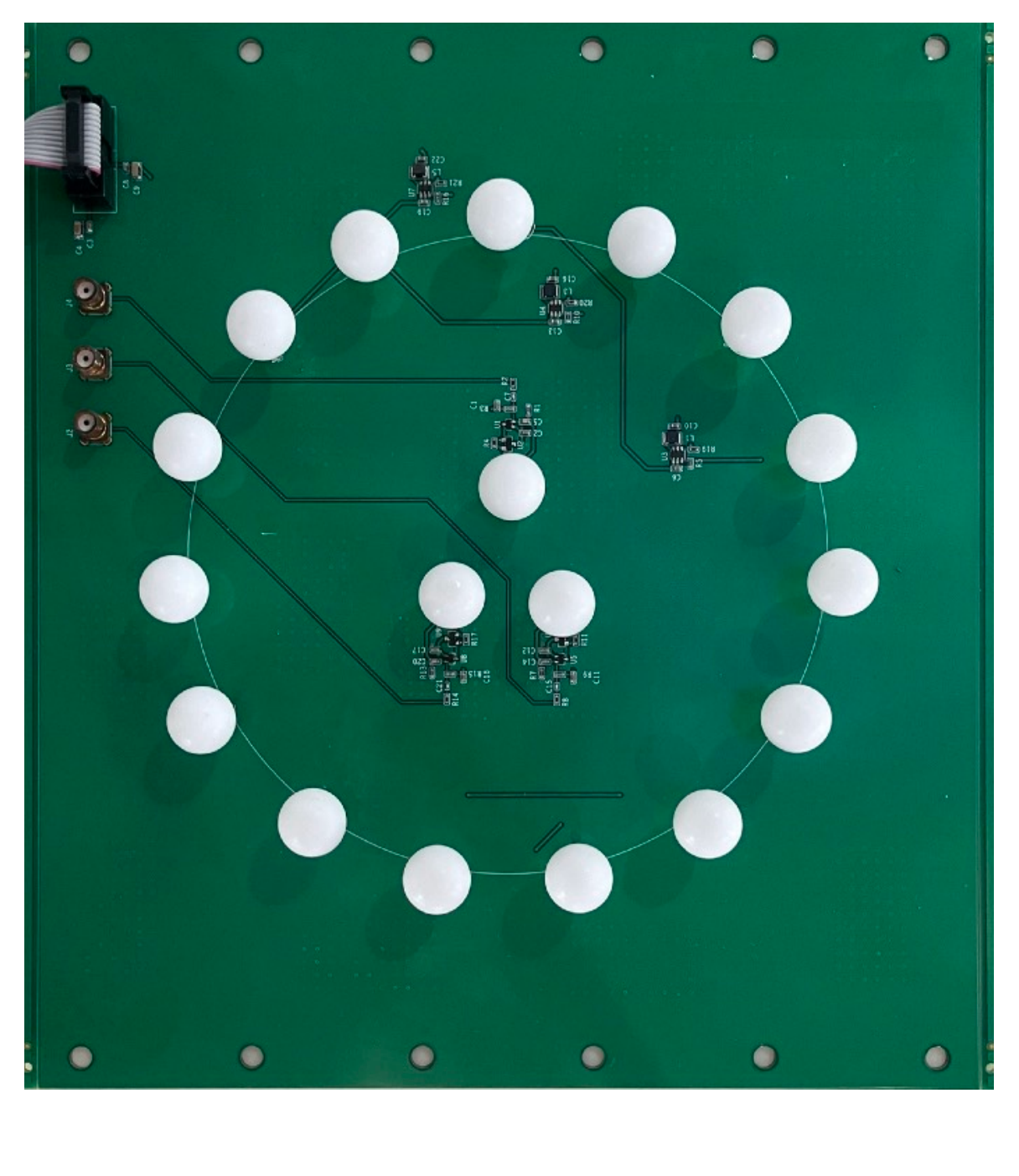}
    \caption{\label{fig_LED_layout_b}LED board used in T-REX. }
\end{subfigure}
\caption{\label{fig_LED_layout} LED layout. The center of the board comprises three pulsing LEDs emitting at wavelengths of $405~\mathrm{nm}$, $450~\mathrm{nm}$, and $525~\mathrm{nm}$. The outer ring consists of fifteen steady LEDs, arranged uniformly.
}
\end{figure}

The layout of the LEDs is shown in Figure~\ref{fig_LED_layout_a}, and the LED board used in T-REX is presented in Figure~\ref{fig_LED_layout_b}.
All LEDs are covered with Teflon balls. The pulsing light source consists of three pulsing LEDs arranged in a triangular shape at the center of the LED board. To enable the camera to capture the light source at a distance of tens of meters, we use five identical LEDs for each wavelength arranged equidistantly on a ring with a radius of $7.5~\mathrm{cm}$ for the steady light source. Fifteen steady LEDs with three wavelengths are alternately arranged on the ring to make the light field more uniform.

\subsection{LED selection}

LEDs with multiple wavelengths are crucial components of the light source.
The wavelengths of LEDs should cover the major wavelengths of Cherenkov radiation in water and fall within the PMT's sensitive detection range.
Previous measurements of ANTARES \cite{ANTARES:2005}, KM3NeT~\cite{KM3NeT:2011ioj}, P-ONE STRAW-a \cite{Bailly:2021dxn}, and Baikal-GVD \cite{baikal_neutrino2022} have shown that the attenuation length is relatively long in the wavelength range of $400~\mathrm{nm}$ to $500~\mathrm{nm}$, with a peak at $\sim 460~\mathrm{nm}$.
Therefore, we preselected and tested multiple sets of LEDs with wavelengths ranging from $380~\mathrm{nm}$ to $590~\mathrm{nm}$.

\begin{table}[htbp]
\centering
\caption{\label{tab_1} The selected LEDs for T-REX. Given are the LED names, emitted colors, wavelengths with the measured FWHM spectral ranges, and the operation mode. }
\smallskip 
\begin{tabular}{cccc}
\hline
LED & Color & Wavelength[nm] & Mode\\
\hline
UV3TZ-405-30 & violet & $405\pm 7$ & pulsing/steady\\
STC450-5C10 & blue & $450\pm 11$ & pulsing\\ 
YY-B5012UC & blue  & $460\pm 11$ & steady\\  
WP710A10LZGCK  & green & $525\pm 17$ & pulsing/steady\\
\hline
\end{tabular}
\end{table}

\begin{figure}[htb]
\captionsetup[subfigure]{justification=Centering}
\begin{subfigure}[t]{0.48\textwidth}
    \includegraphics[width=\linewidth]{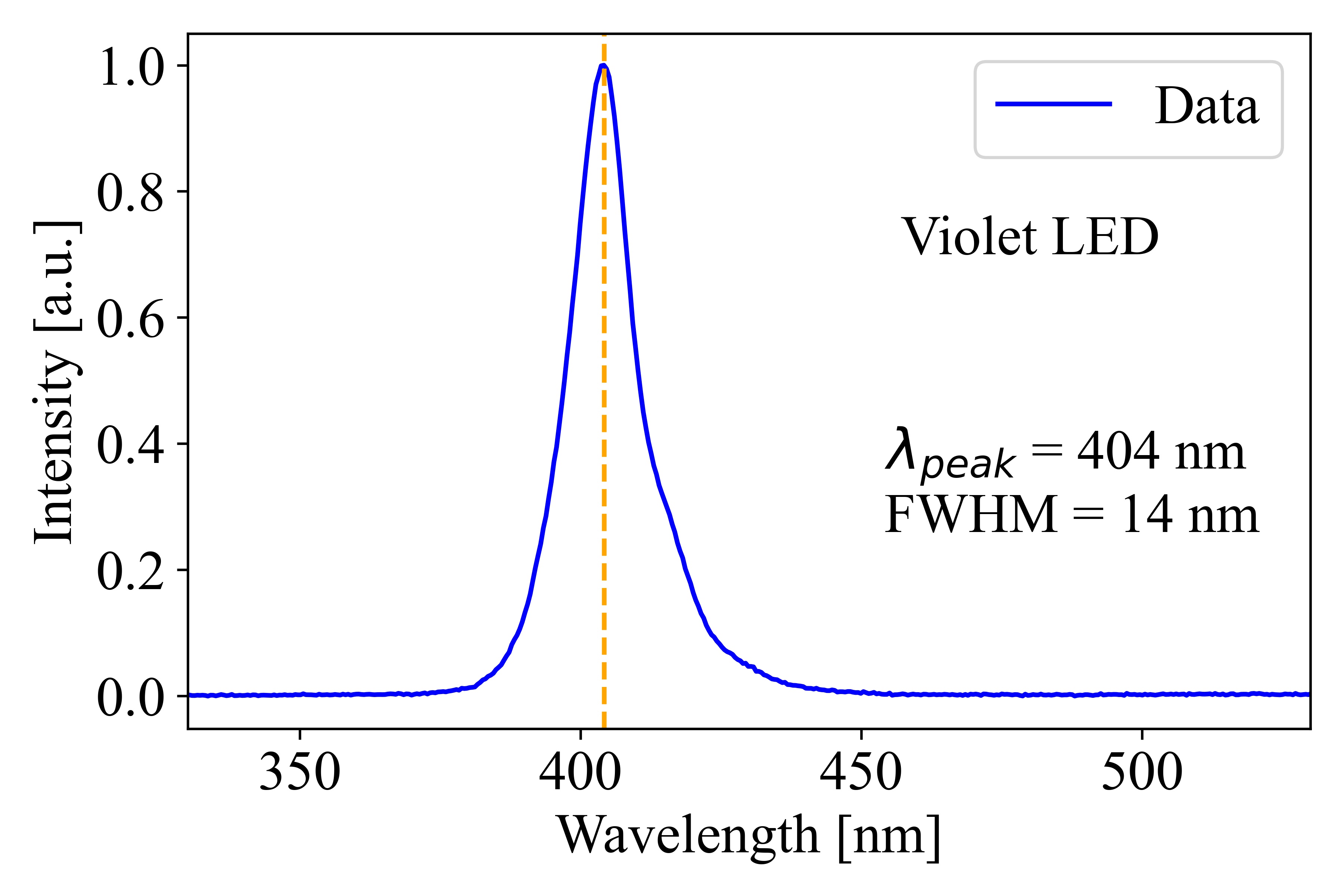}
    \caption{ $405~\mathrm{nm}$ LED spectrum.}
\end{subfigure}\hspace{\fill}
\begin{subfigure}[t]{0.48\textwidth}
    \includegraphics[width=\linewidth]{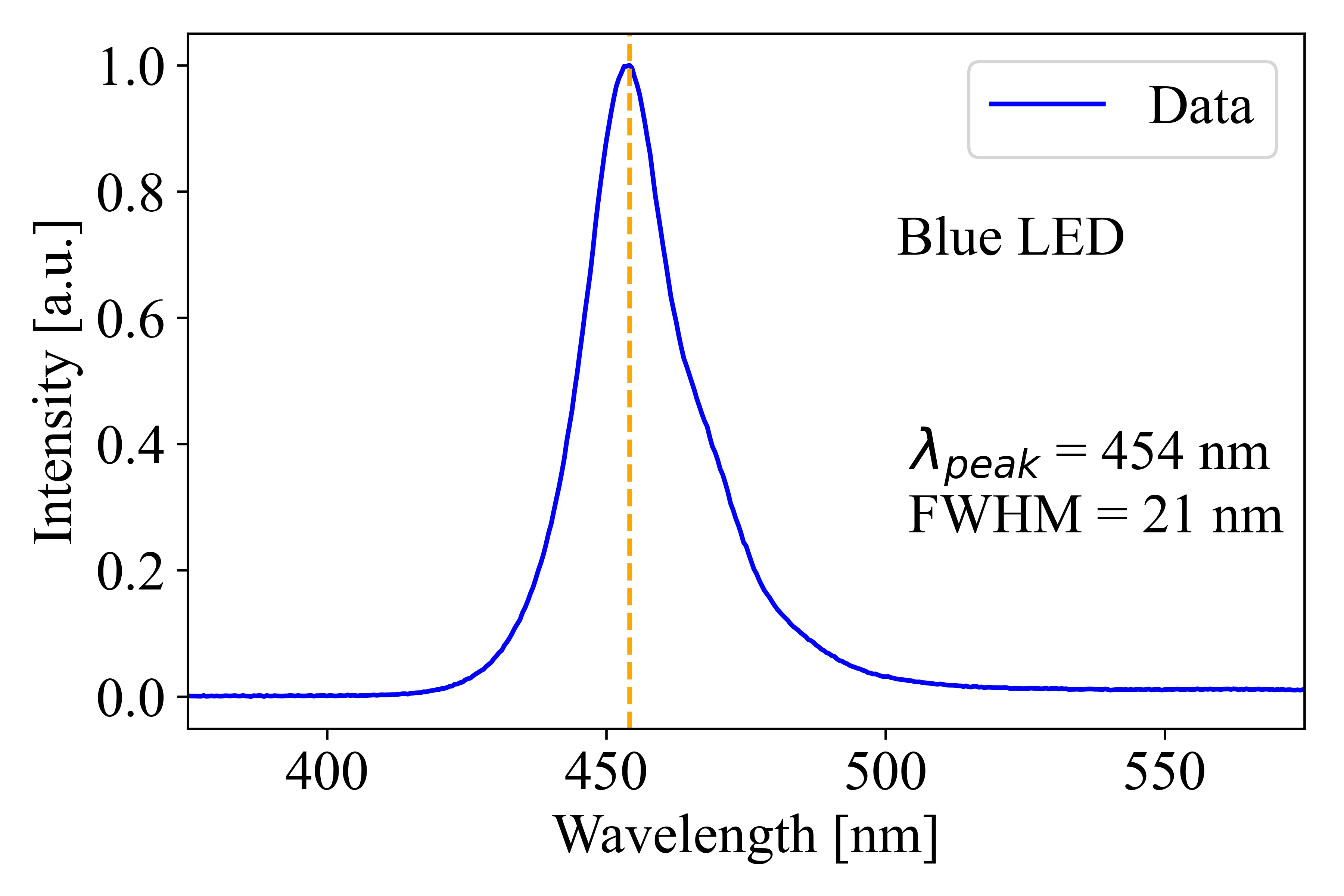}
    \caption{ $450~\mathrm{nm}$ LED spectrum.}
\end{subfigure}
\begin{subfigure}[t]{0.48\textwidth}
    \includegraphics[width=\linewidth]{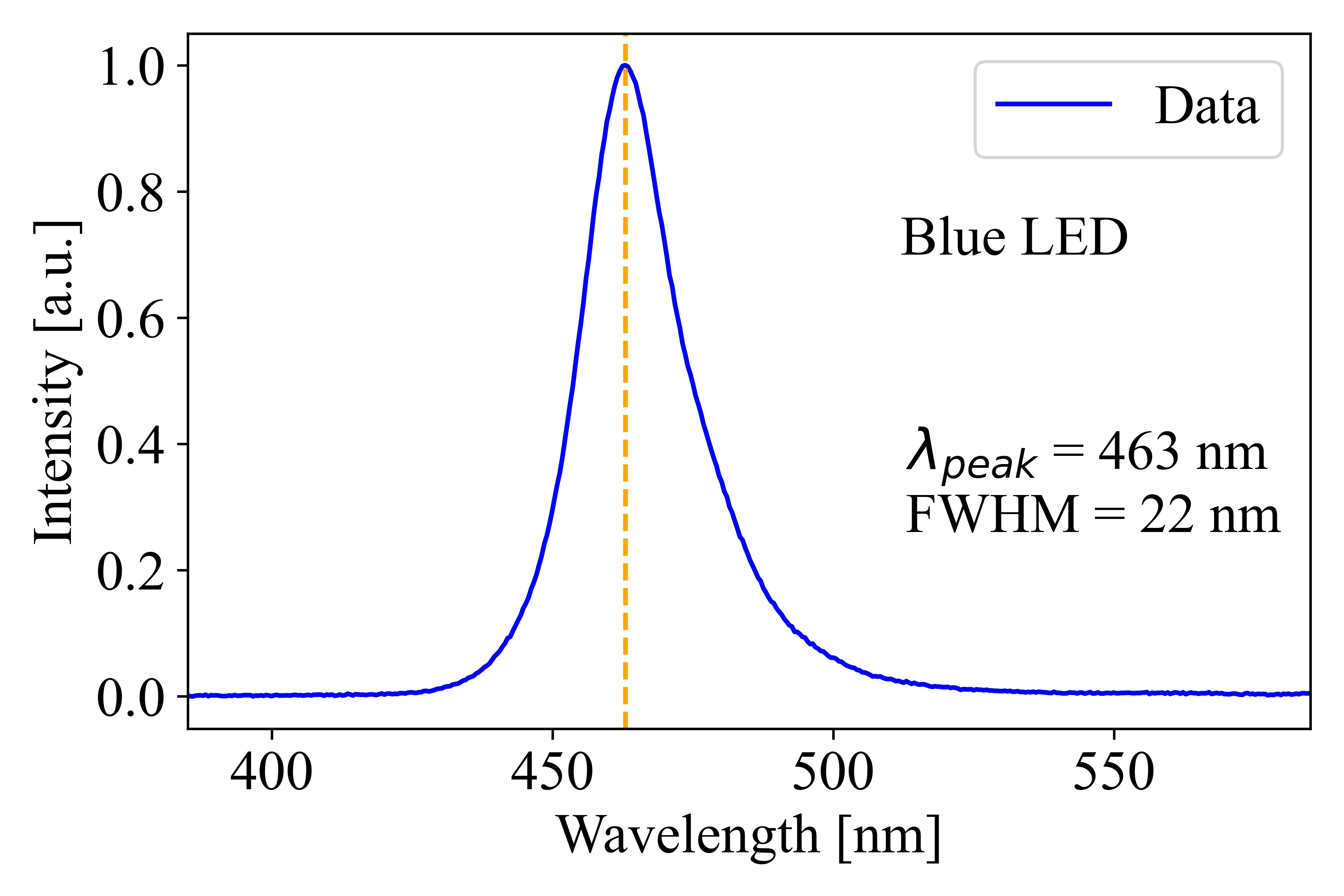}
    \caption{ $460~\mathrm{nm}$ LED spectrum.}
\end{subfigure}\hspace{\fill}
\begin{subfigure}[t]{0.48\textwidth}
    \includegraphics[width=\linewidth]{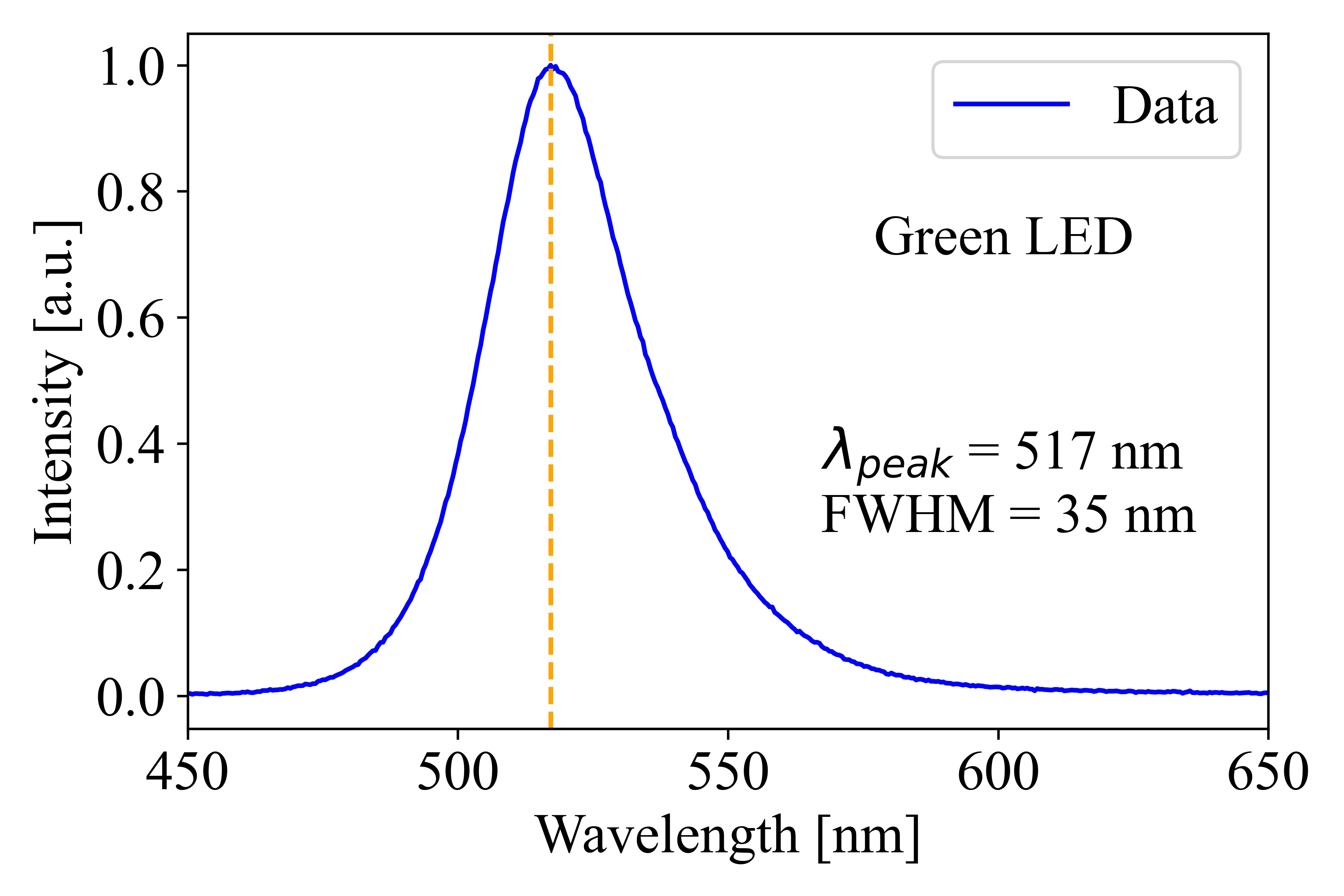}
    \caption{ $525~\mathrm{nm}$ LED spectrum.}
\end{subfigure}
\caption{\label{fig_led_spectrum} 
Spectra of LEDs used in T-REX. The spectra were measured at steady mode. LEDs were driven by a constant current of $20~\mathrm{mA}$. The intensity is in arbitrary units. The peak wavelength and FWHM of each spectrum are shown in the figure. The measured peak wavelengths exhibit slight deviations from the nominal manufacturer wavelengths, as presented in Table~\ref{tab_1}.
}
\end{figure}

The selection of LEDs primarily depends on their time profile and luminous intensity. Firstly, we measured the pulse time profiles of different LEDs and preselected LEDs with a narrower time spread width, as discussed in Section~\ref{led_selection_timeprofile}. Subsequently, we tested the preselected LEDs for their luminous intensity in both pulsing and steady modes, as described in Section~\ref{led_selection_brightness}. Based on these measurements, we selected four LEDs for use in T-REX. Table~\ref{tab_1} provides a summary of all the LEDs used in T-REX, while Figure~\ref{fig_led_spectrum} shows the light spectra of the selected LEDs measured by a spectrometer. 
Despite the measured wavelengths of the LEDs being slightly deviated from their typical values, they are still referred to by their nominal wavelengths of $405~\mathrm{nm}$, $450~\mathrm{nm}$, $460~\mathrm{nm}$, and $525~\mathrm{nm}$.

\subsubsection{Time profile}
\label{led_selection_timeprofile}

\begin{figure}[htb]
\centering 
\includegraphics[scale=0.4]{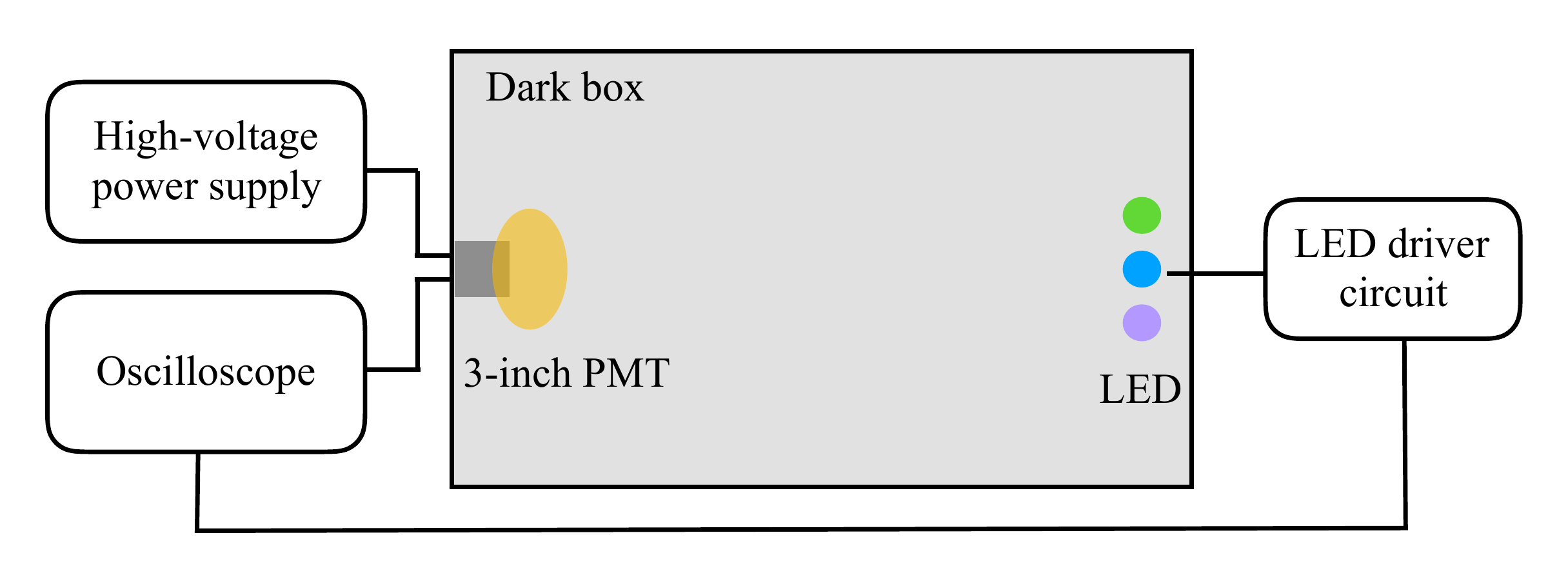}
\caption{\label{fig_time_spread_setup} Sketch of the laboratory setup for measurements of the pulse time profile of LEDs.}
\end{figure}

\begin{figure}[htb]
\centering 
\includegraphics[width=0.48\linewidth]{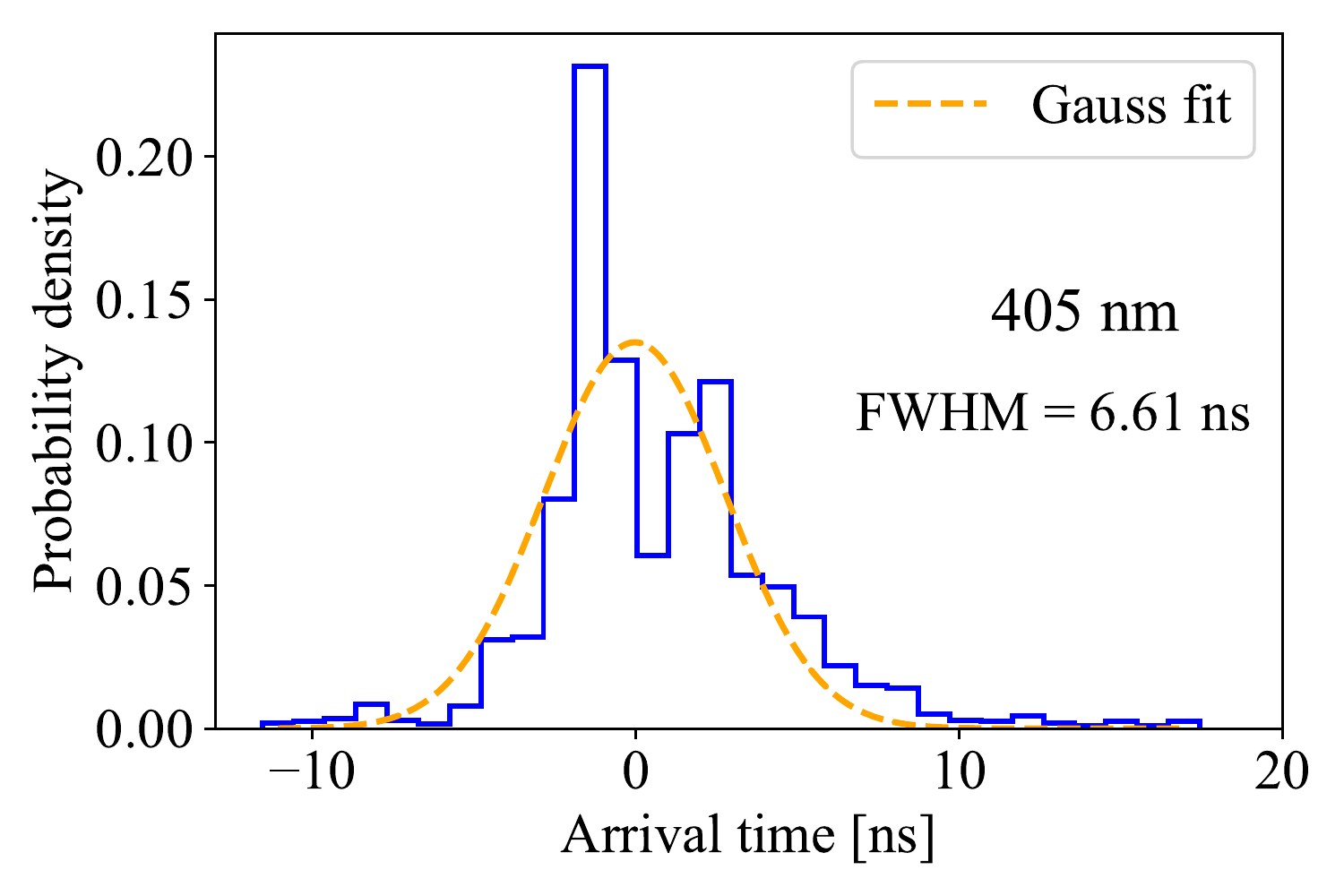}
\quad
\includegraphics[width=0.48\linewidth]{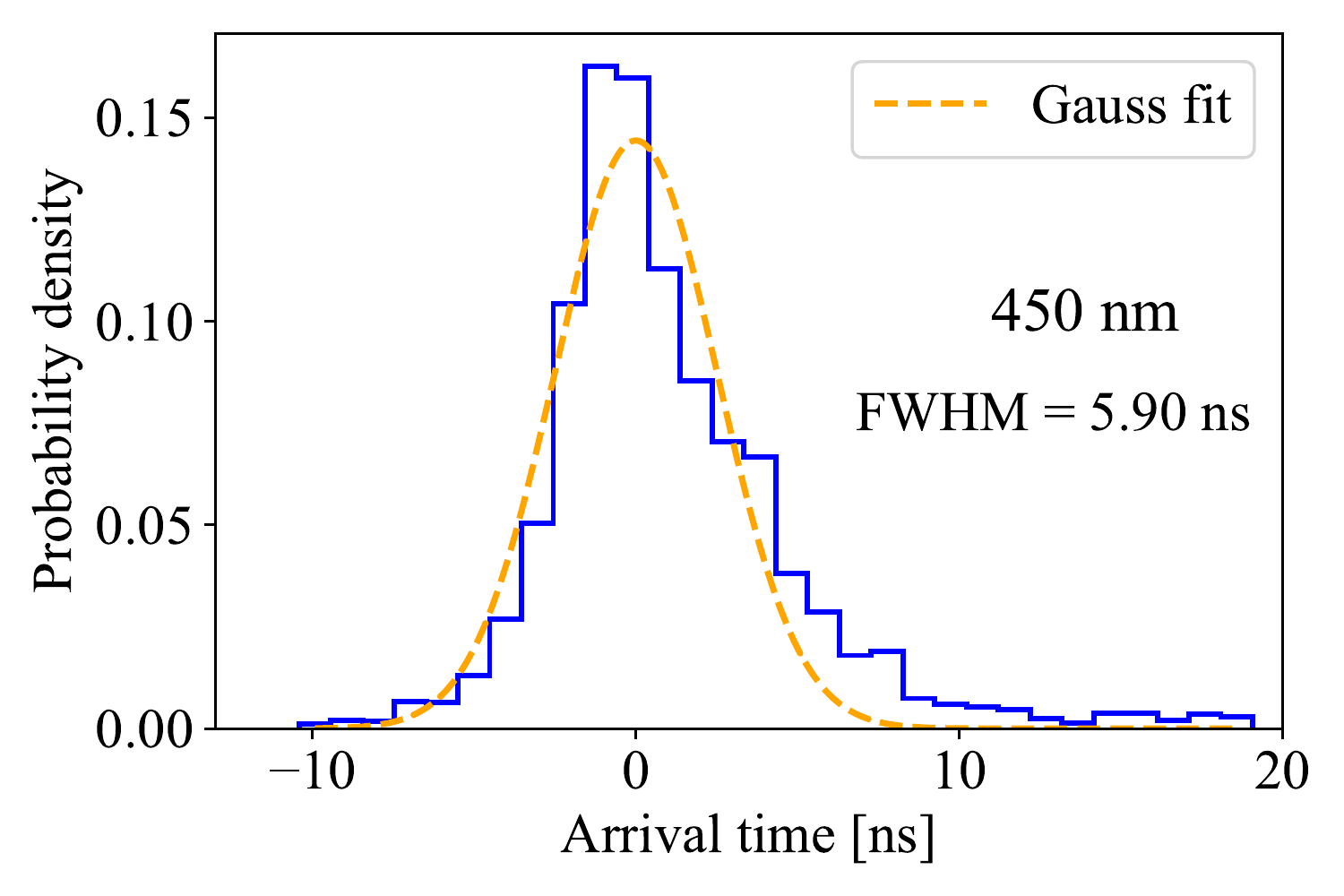}
\includegraphics[width=0.48\linewidth]{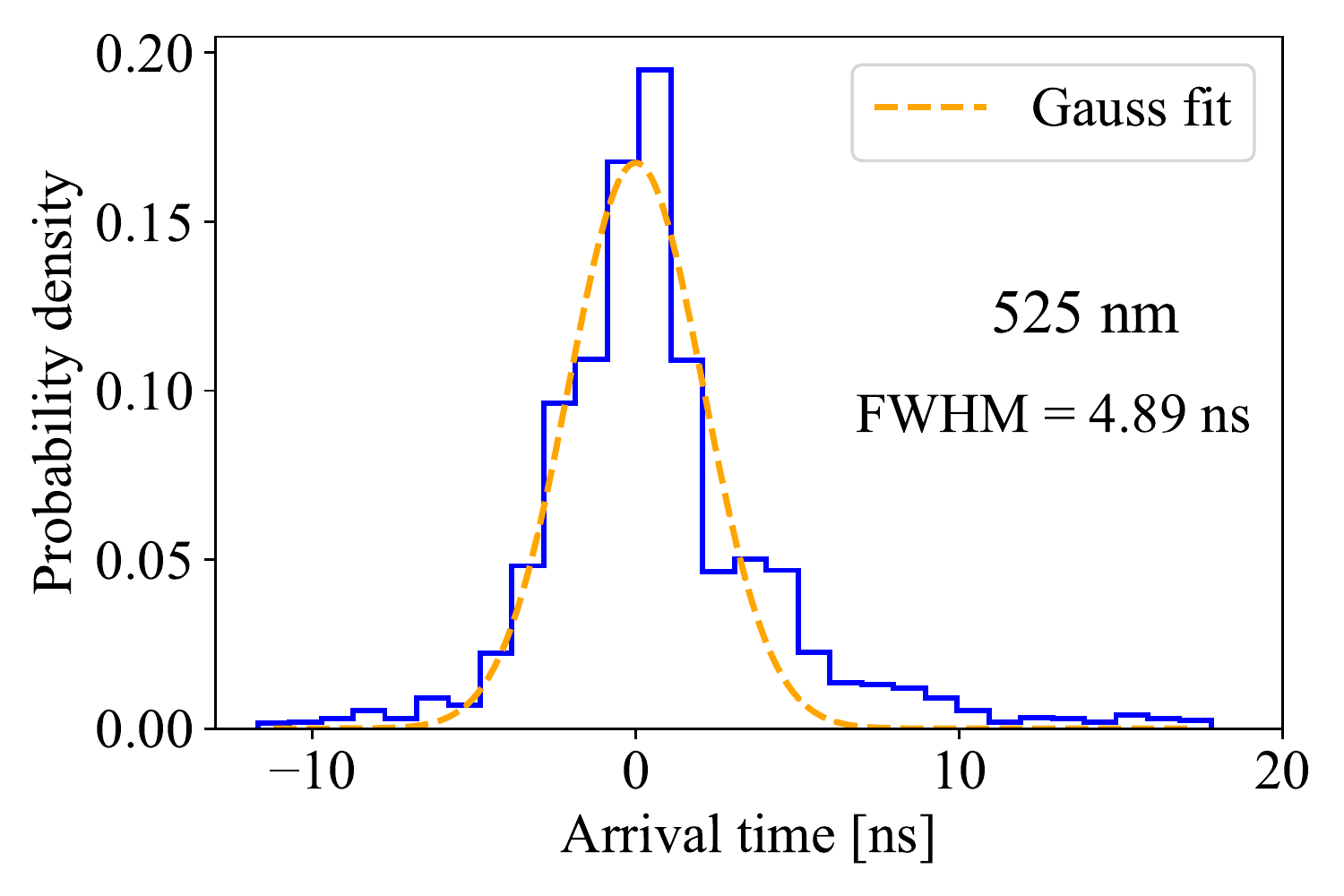}
\caption{\label{fig_time_spread} Time profiles of  $405~\mathrm{nm}$, $450~\mathrm{nm}$, and $525~\mathrm{nm}$ LEDs. The Gaussian fitting to the data is shown in orange. The FWHM of each LED's time profile, which includes a $\sim 4~\mathrm{ns}$ TTS of the PMT, is shown in each panel. }
\end{figure}

The laboratory setup for measuring pulse time profiles of LEDs is depicted in Figure~\ref{fig_time_spread_setup}. A 3-inch PMT and LEDs covered with Teflon balls were placed opposite to each other inside a dark box. 
To obtain SPE signals reliably, the supply voltage of LEDs was adjusted to achieve an average signal expectation of $0.1$ photoelectrons for the PMT.
The PMT was operated at a high voltage of $\sim 1375~\mathrm{V}$ and connected to an oscilloscope for recording the waveforms.
Each LED was tested separately using the driver circuit employed in T-REX. The arrival times of photons were obtained by analyzing the waveforms of the SPE. The time profiles of selected pulsing LEDs are shown in Figure~\ref{fig_time_spread}, which are fitted with a Gaussian distribution. The full width at half maximum (FWHM) of the time profiles that incorporate a $\sim 4~\mathrm{ns}$ transit time spread (TTS) of the PMT \textcolor{blue}{\cite{Yudi:2022}} are $6.61~\mathrm{ns}$ ($405~\mathrm{nm}$), $5.90~\mathrm{ns}$ ($450~\mathrm{nm}$), and $4.89~\mathrm{ns}$ ($525~\mathrm{nm}$), respectively.

\subsubsection{Luminous intensity}
\label{led_selection_brightness}

For the pulsing LEDs, we measured the luminous intensity of preselected LEDs with the laboratory setup shown in Figure~\ref{fig_led_PMT_calbiration}. 
The pulsing LED luminous intensity measurement experiment was conducted in a $3.5~\mathrm{m}$ long customized dark water tank. The water tank was covered with black cloths to prevent light reflection. A 3-inch PMT was used to calibrate the pulsing LEDs in the light source, with the light source and the PMT positioned opposite to each other at a distance of $2~\mathrm{m}$. An attenuator with a transmittance of 0.17\% at $450~\mathrm{nm}$ was placed in front of the PMT to simulate the attenuation of light in seawater at tens of meters distance. The attenuation rate, which depends on the number of attenuators and the PMT-light source distance, was pretested to account for unknown seawater attenuation length and potential fluctuations. The light source and the PMT were connected to the computer via a coarse wavelength division multiplexer (CWDM).
We measured the signal occurrence probability and signal average charge integral over a voltage range of $8~\mathrm{V}$ to $25~\mathrm{V}$. The $405~\mathrm{nm}$, $450~\mathrm{nm}$, and $525~\mathrm{nm}$ LEDs with the highest product of the signal occurrence probability and signal average charge integral, corresponding to $1.7$, $1.1$, and $1.3$ photoelectrons at a voltage of 15V, respectively, were selected as the pulsing LEDs. 

For the steady LEDs, we measured the luminous intensity of the steady light source at a constant current of $20~\mathrm{mA}$ using a photometer. LEDs with wavelengths of $405~\mathrm{nm}$, $460~\mathrm{nm}$, and $525~\mathrm{nm}$ were selected as steady LEDs, corresponding to measured luminous intensities of $35$, $42$, and $5~\mathrm{lux}$, respectively.

\subsection{Pulsing LED calibration}
\label{pulsing_LED_calibration}

\begin{figure}[htbp]
\centering 
\includegraphics[scale=0.35]{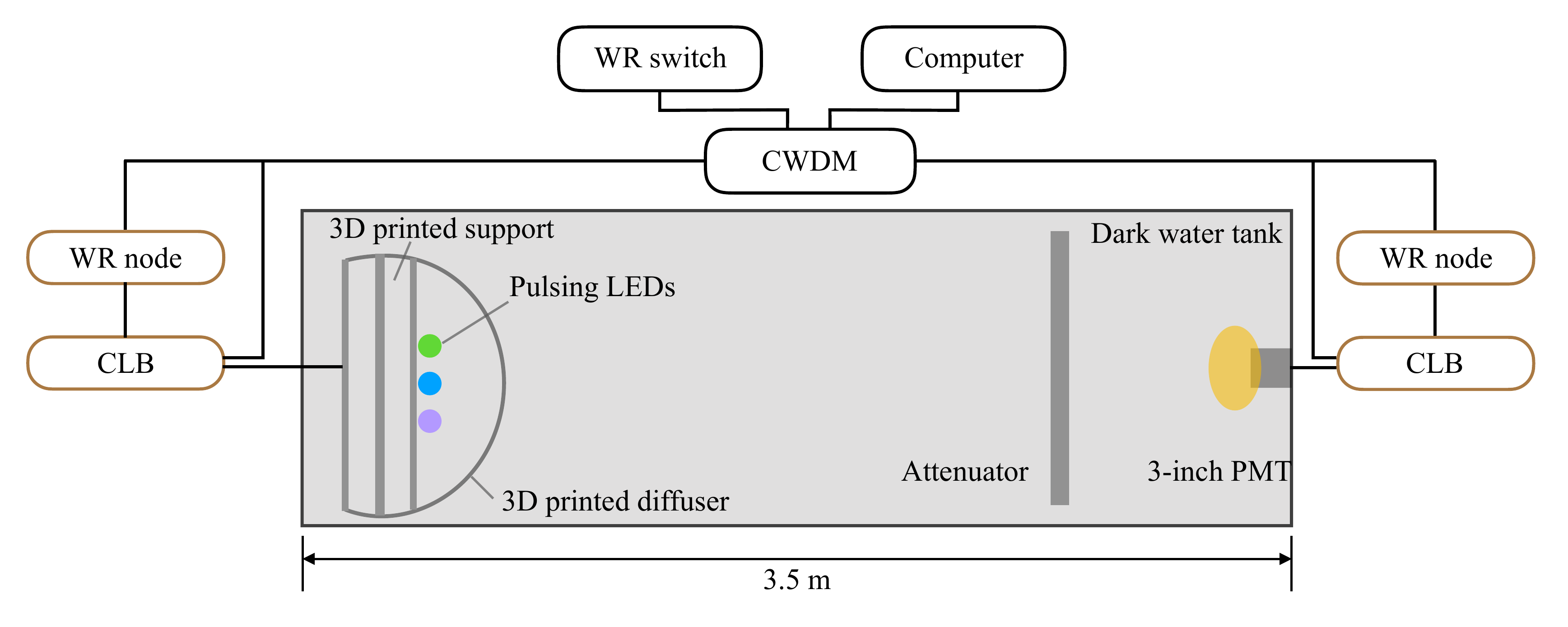}
\caption{\label{fig_led_PMT_calbiration} Sketch of the laboratory setup for measurements of the average number of photoelectrons received by the PMT as a function of the supply voltage of the pulsing LEDs. The pulsing LEDs were covered with diffusers. The electronic boards were connected to the computer and WR switch via a CWDM. }
\end{figure}

To verify that PMTs in receiver modules can receive SPE signals from the pulsing light source and to account for potential variations in deep-sea conditions, we calibrated the average number of photoelectrons received by the PMT as a function of the supply voltage of the pulsing LEDs.
The laboratory setup for the pulsing LED-PMT calibration experiment is shown in Figure~\ref{fig_led_PMT_calbiration}.
The supply voltage of the pulsing LED was controlled by the slow control system, which automatically scanned the voltage from $8~\mathrm{V}$ to $25~\mathrm{V}$ with a step of $1~\mathrm{V}$. 

\begin{figure}[htb]
\centering 
\includegraphics[width=0.48\linewidth]{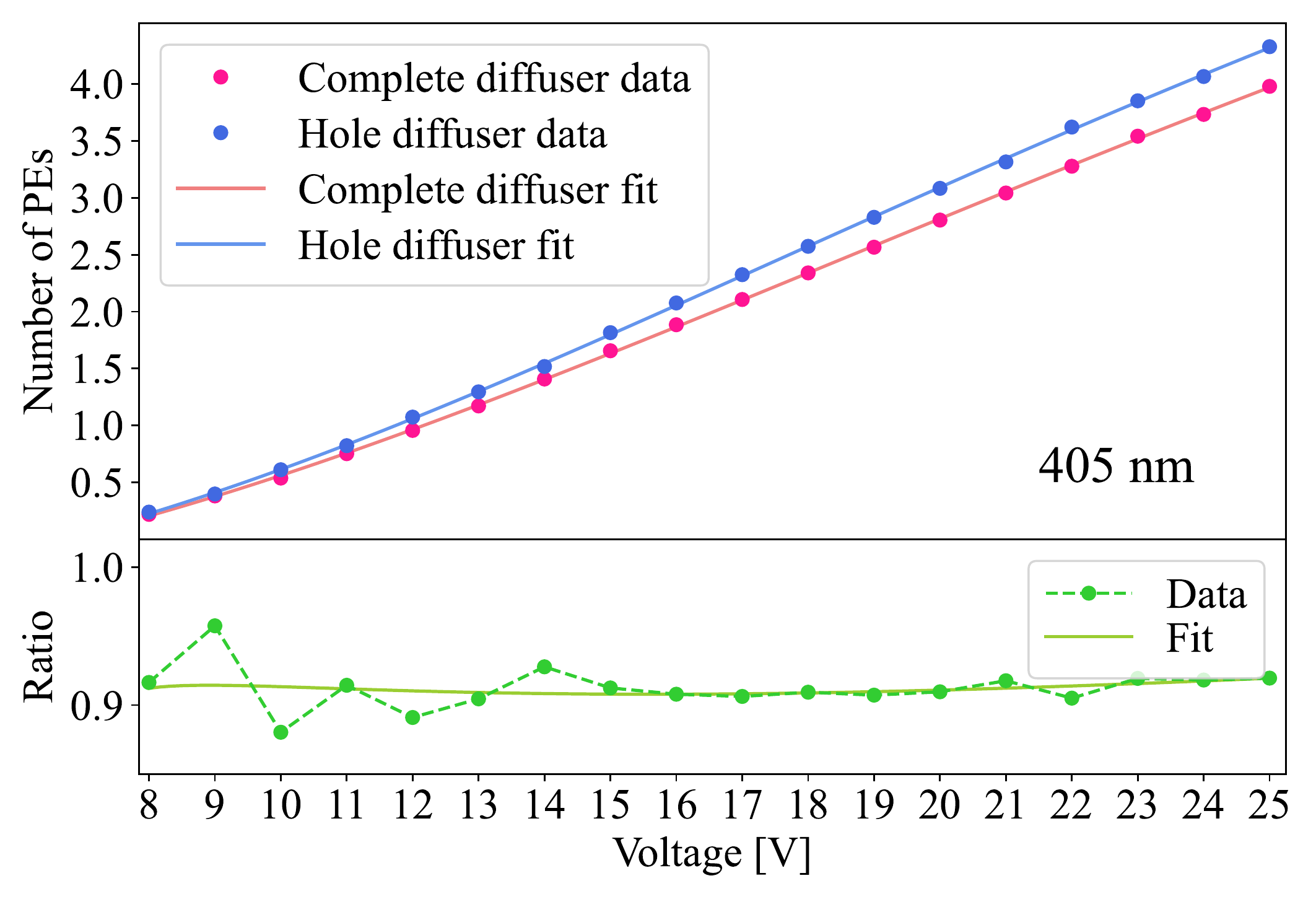}
\quad
\includegraphics[width=0.48\linewidth]{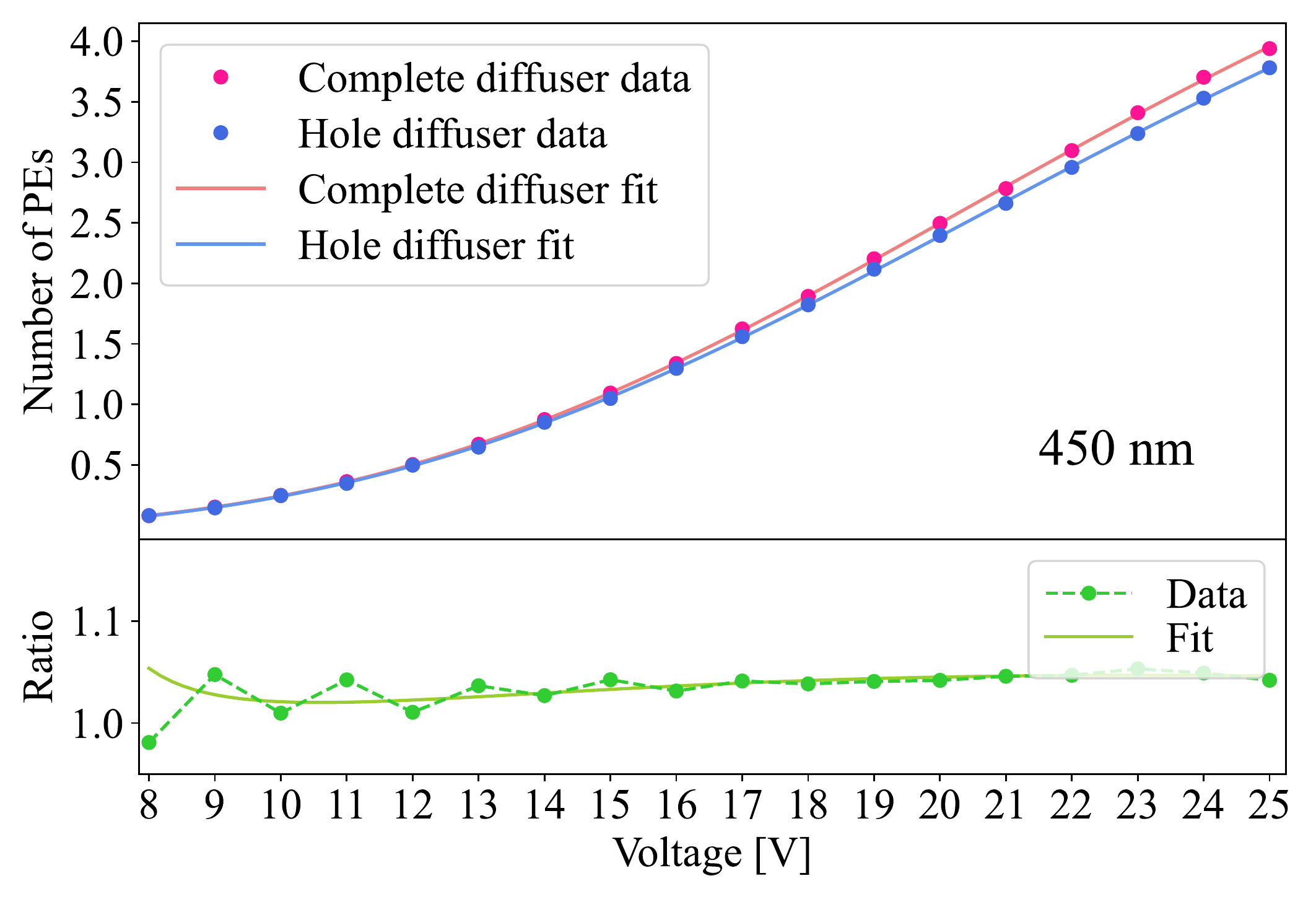}
\includegraphics[width=0.48\linewidth]{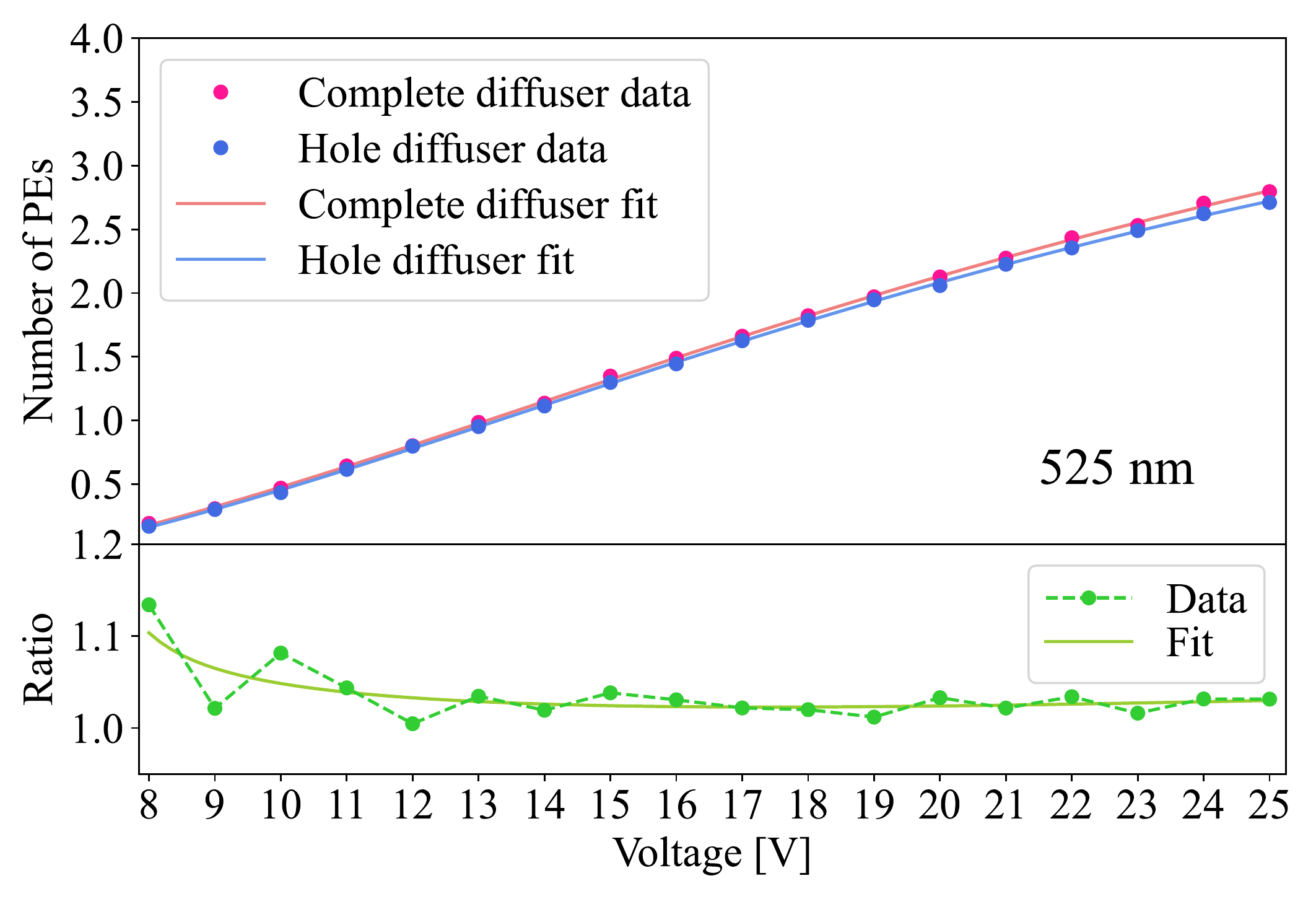}
\caption{\label{fig_PMT_LED_Voltage} The average number of photoelectrons received by the PMT as a function of the supply voltage of the pulsing LED. The lower panel of each figure shows the ratio of the average number of photoelectrons received by the PMT of the complete hemisphere over the hole hemisphere. }
\end{figure}

Figure~\ref{fig_PMT_LED_Voltage} shows the average number of photoelectrons received by the PMT as a function of the supply voltage of LEDs at three different wavelengths. 
The results demonstrate a relatively good linear relationship between the number of photoelectrons received by the PMT and the supply voltage of the pulsing LED.
The luminous intensity of the pulsing LED is sufficiently bright for the deep-sea experiment since we have added an attenuator to simulate the attenuation of seawater. 
The PMT can receive SPE signals from the pulsing light source within this supply voltage range.
Although the number of photoelectrons received by the PMT from the complete hemisphere and the hole hemisphere of the light source differs slightly, the brightness ratio caused by this difference was well-calibrated, as described in Section \ref{sec:pulsing_light_source_calibation}.

\subsection{Steady LED calibration}
\begin{figure}[htbp]
\centering 
\includegraphics[scale=0.12]{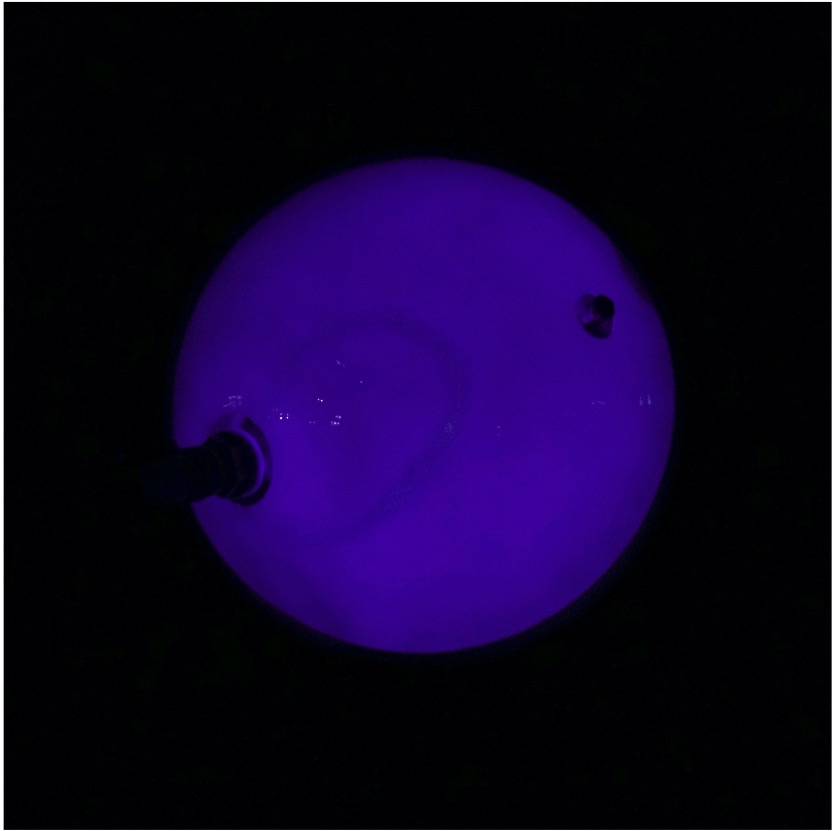}
\quad
\includegraphics[scale=0.12]{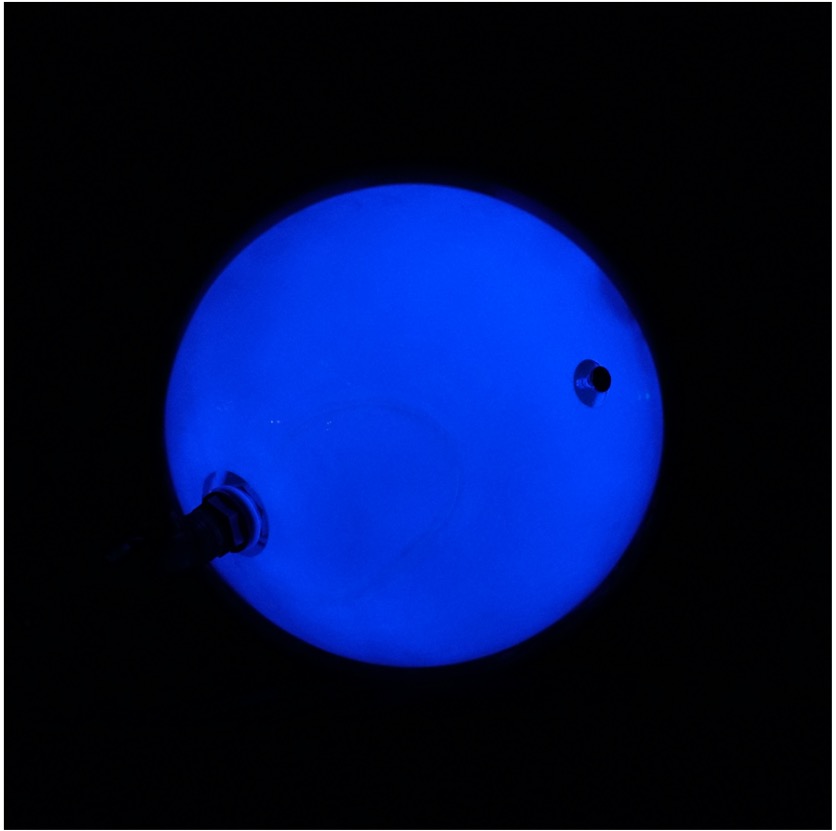}
\quad
\includegraphics[scale=0.12]{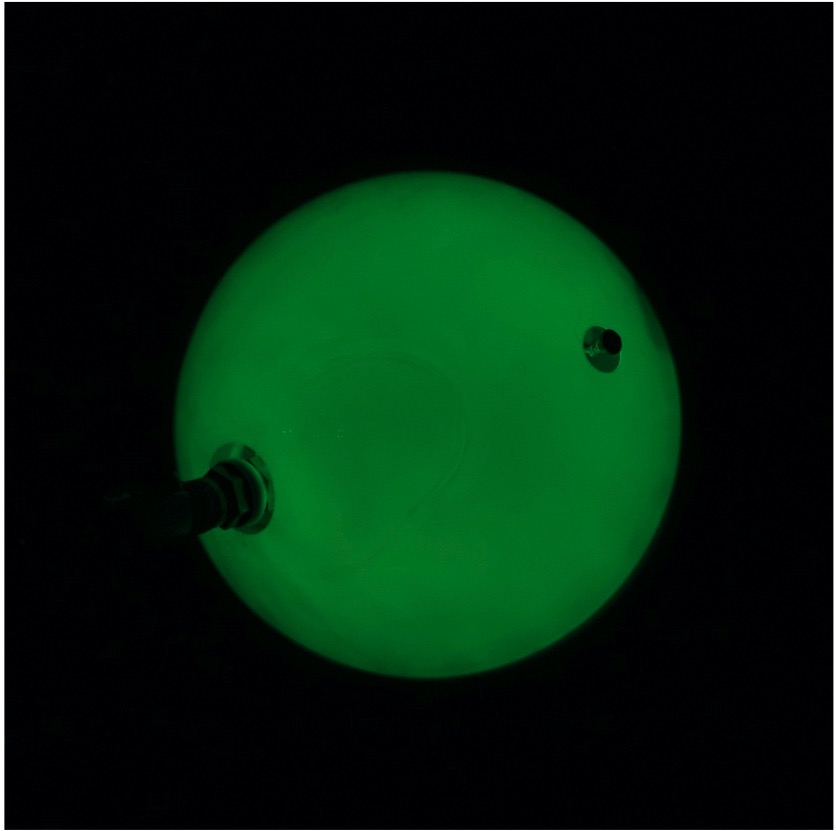}
\includegraphics[scale=0.18]{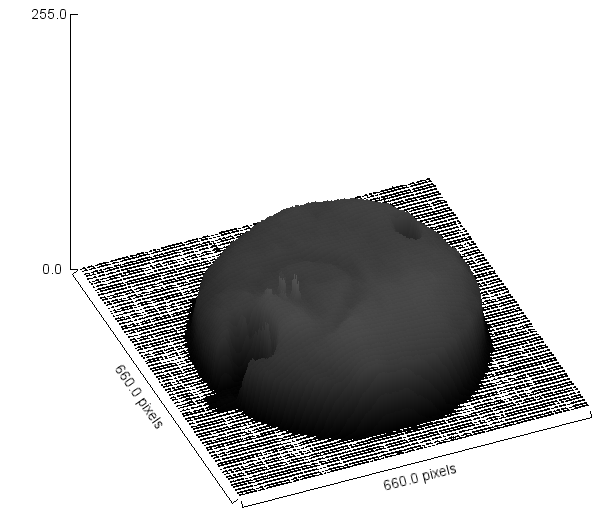}
\includegraphics[scale=0.18]{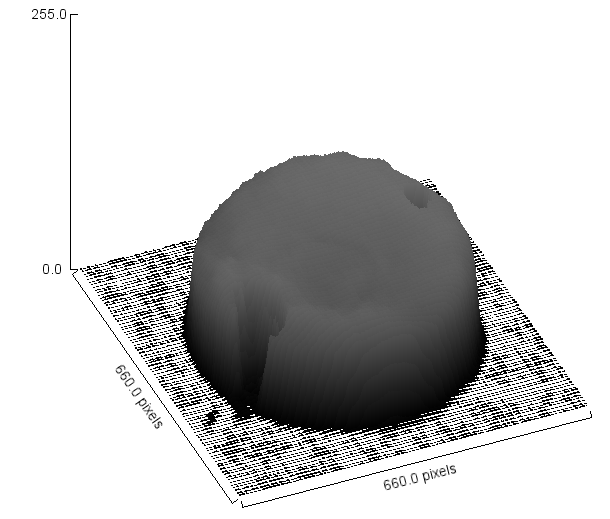}
\includegraphics[scale=0.18]{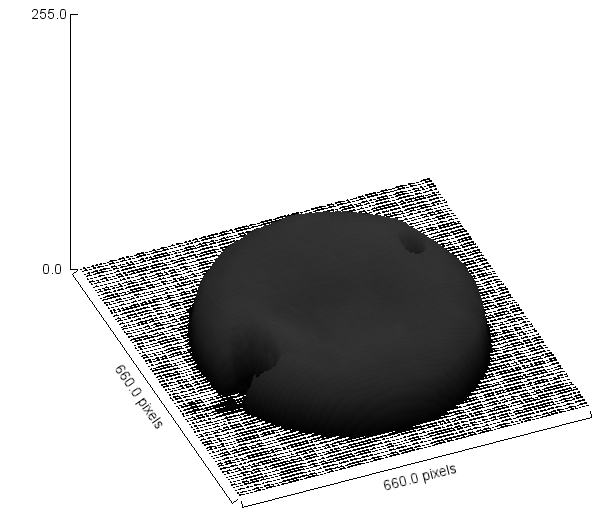}
\caption{\label{fig_steady_light_source} Images (upper panel) and surface plots (lower panel) of the $405~\mathrm{nm}$, $460~\mathrm{nm}$, and $525~\mathrm{nm}$ steady light source. The two black shadows in each image are the vacuum port and penetrator. In the lower panel, the z-axis represents the gray value, while the x and y-axes correspond to the pixel coordinates.}
\end{figure}

The steady light source of each wavelength consists of five identical LEDs connected in series and driven by a constant current of $20~\mathrm{mA}$.
To validate the operational state of the steady LEDs under this drive current, we used a color camera to capture images at room temperature. The upper panel of Figure~\ref{fig_steady_light_source} shows the images of the steady light source at three wavelengths, while the lower panel shows a surface plot of the light source, with the z-axis representing the gray value of each pixel. The steady light source performed well at room temperature.

\section{Light source calibration at low temperature}
\label{sec:low temperature calibration}

The temperature of \textit{in-situ} deep-sea water is relatively low and stays stable at $\sim 2^{\circ}C$.
To study the light source responses at its designated working temperature, we conducted a series of calibration experiments in a temperature-controlled dark room.
The dark room is $10~\mathrm{m}$ long and $2~\mathrm{m}$ wide. We verified that the light source exhibited good luminescent stability, with a brightness ratio variation of less than $2\%$ between room temperature ($18.9\pm 0.5^{\circ}C$) and low temperature ($2.1\pm 0.5^{\circ}C$).

Both the PMT and camera systems use a relative measurement strategy to decipher the optical properties.
We conducted separate calibrations of the brightness ratio for the steady and pulsing light source. Additionally, we calibrated the emission profile of the light source in the steady mode to investigate its isotropy.

\subsection{Steady light source calibration}

\begin{figure}[htbp]
\centering 
\includegraphics[scale=0.35]{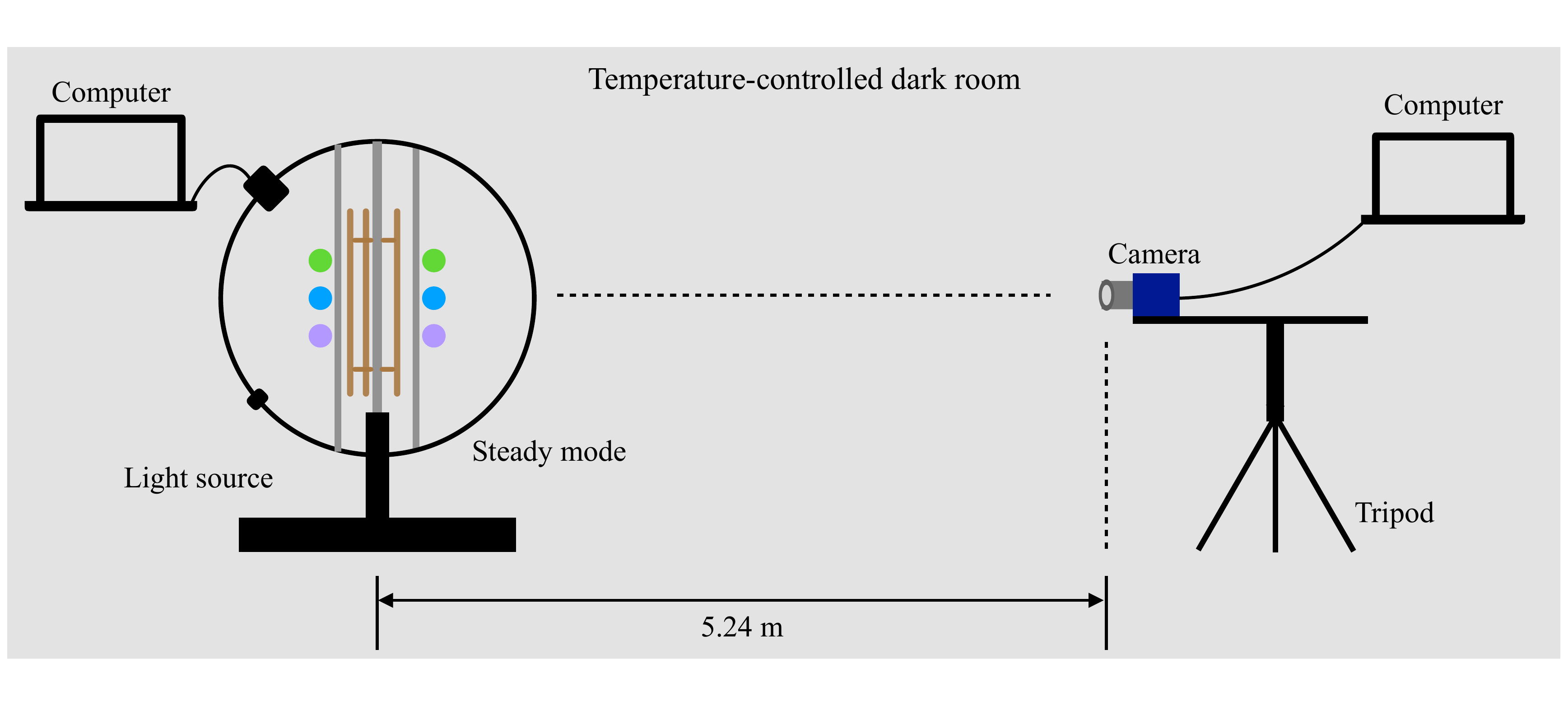}
\caption{\label{fig_camera_setup} Sketch of setup for low-temperature calibration of the steady light source with a camera. The light source was connected to the host computer and controlled by the slow control system. The camera was placed on a tripod and centered on the light source. }
\end{figure}

The laboratory setup for calibrating the brightness ratio and light emission profile is shown in Figure~\ref{fig_camera_setup}. A stand-alone camera was positioned on a tripod opposite the light source at a distance of $5.24\pm 0.01~\mathrm{m}$. The light source was powered by a constant current of $20~\mathrm{mA}$ throughout the calibration process.

\subsubsection{Brightness ratio}

We conducted several groups of calibration experiments with two sets of variables: the temperature ($2.1^{\circ}C$/$18.9^{\circ}C$) and the wavelength of steady LEDs. The camera parameters, including exposure time and gain, were set to be the same as those configured in T-REX. The camera captured images with different set of parameters, and the brightness ratio of the regions of interest was obtained by analyzing the images.

\begin{figure}[htbp]
    \centering
    \includegraphics[width=0.9\textwidth]{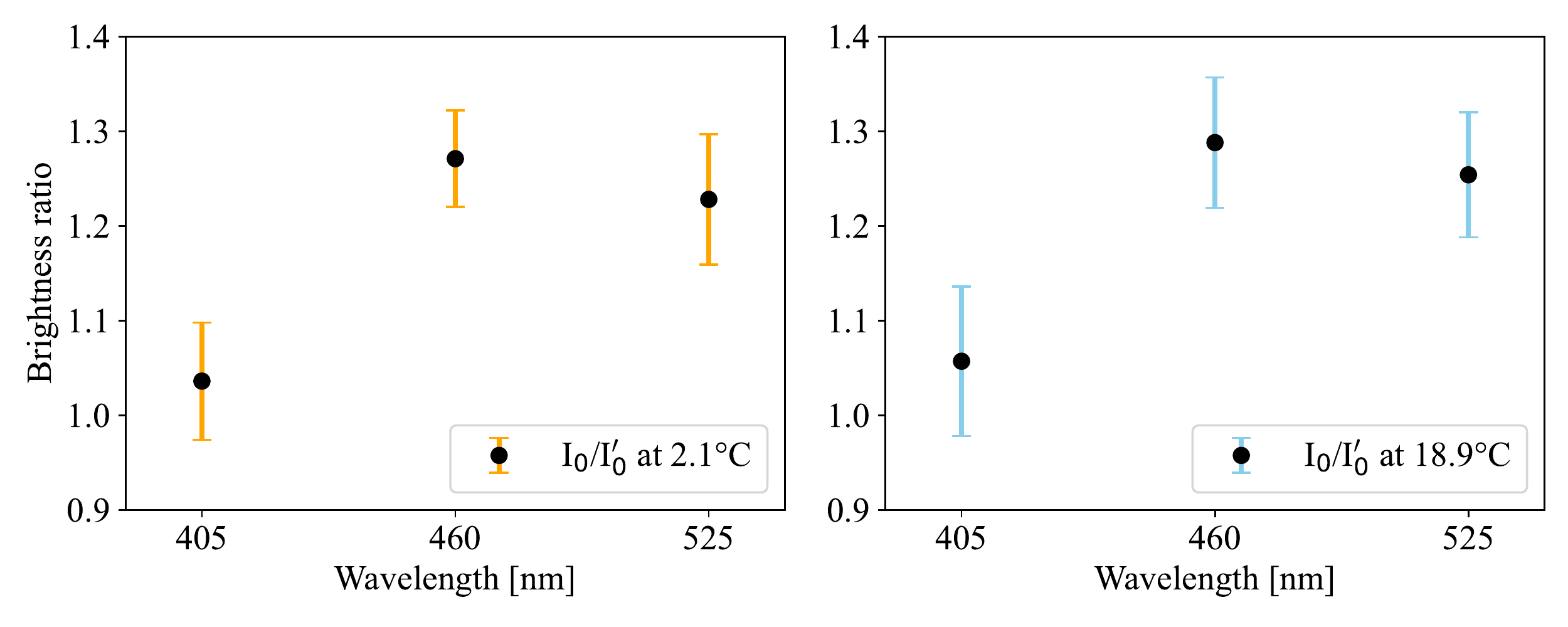}
    \caption{The brightness ratio of the complete hemisphere over the hole hemisphere ($I_0/I_0'$) as a function of wavelength. Left panel: the brightness ratio of the central region at low temperature ($2.1^{\circ}C$). Right panel: the brightness ratio of the central region at room temperature ($18.9^{\circ}C$).}
    \label{fig_steady_light_source_brightness_ratio} 
\end{figure}

We calibrated the brightness ratio of the central region ($I_0/I_0'$), where the central region spans a radius of 40 (21) pixels from the image center for the complete (hole) hemisphere side. The calibrated brightness ratio is used in the analysis of the camera system with the $I_{center}$ method \cite{TianWei:2022}. The brightness ratio for each wavelength at different temperatures is shown in Figure~\ref{fig_steady_light_source_brightness_ratio}. The light source performed stable at different temperatures. The brightness ratio of the complete hemisphere over the hole hemisphere at low temperature was $1.9\%$ ($405~\mathrm{nm}$), $1.6\%$ ($460~\mathrm{nm}$), and $1.6\%$ ($525~\mathrm{nm}$) less than that at room temperature for the central region.

\subsubsection{Light emission profile}
\label{sec:light_emission_profile}

\begin{figure}[htb]
\captionsetup[subfigure]{justification=Centering}
\begin{subfigure}[t]{0.54\textwidth}
    \includegraphics[width=\linewidth]{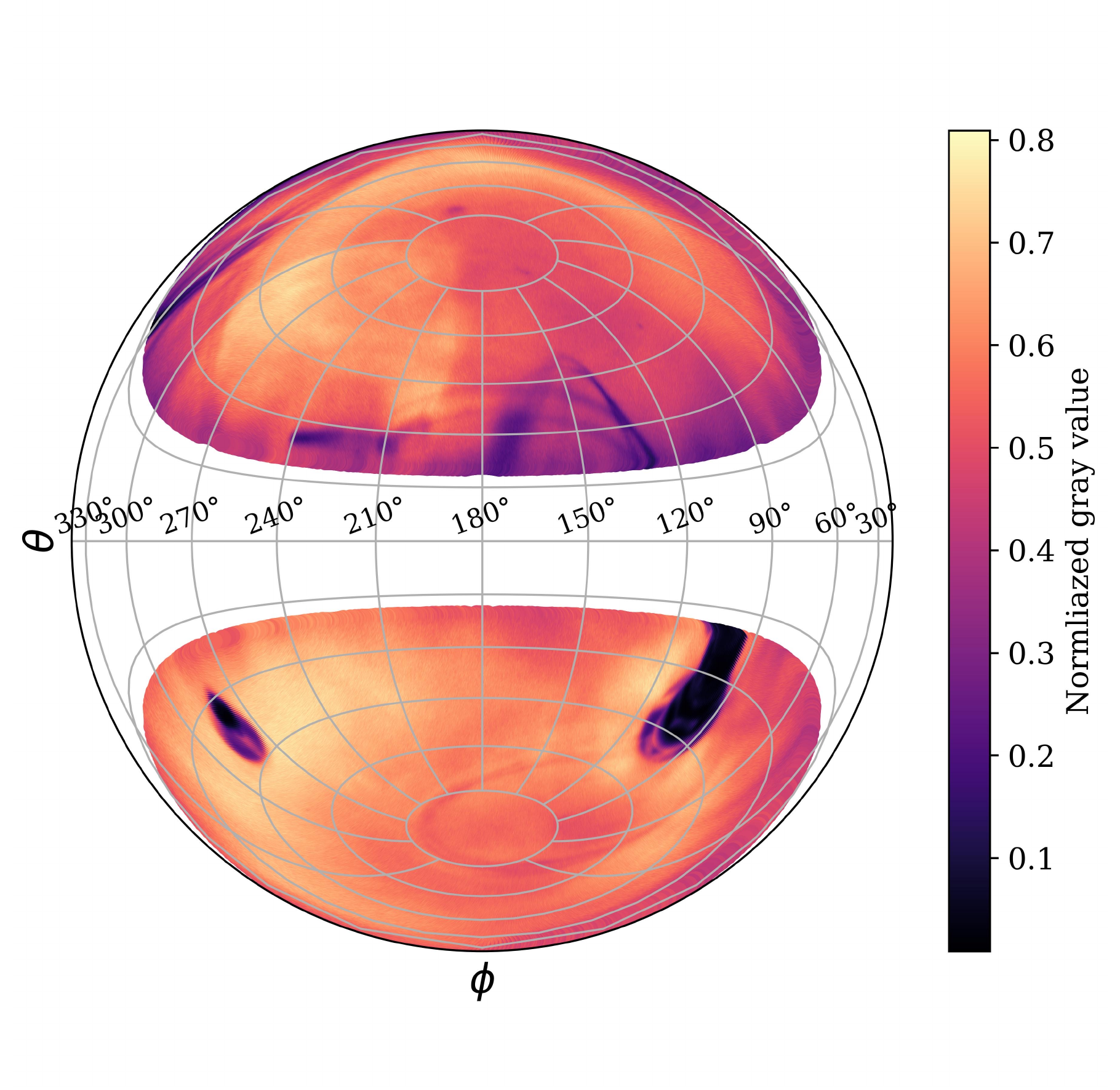}
    \caption{\label{fig_2d_emission_pdf}Emission profile in  Lambert projection. }
\end{subfigure}\hspace{\fill}
\begin{subfigure}[t]{0.44\textwidth}
    \includegraphics[width=\linewidth]{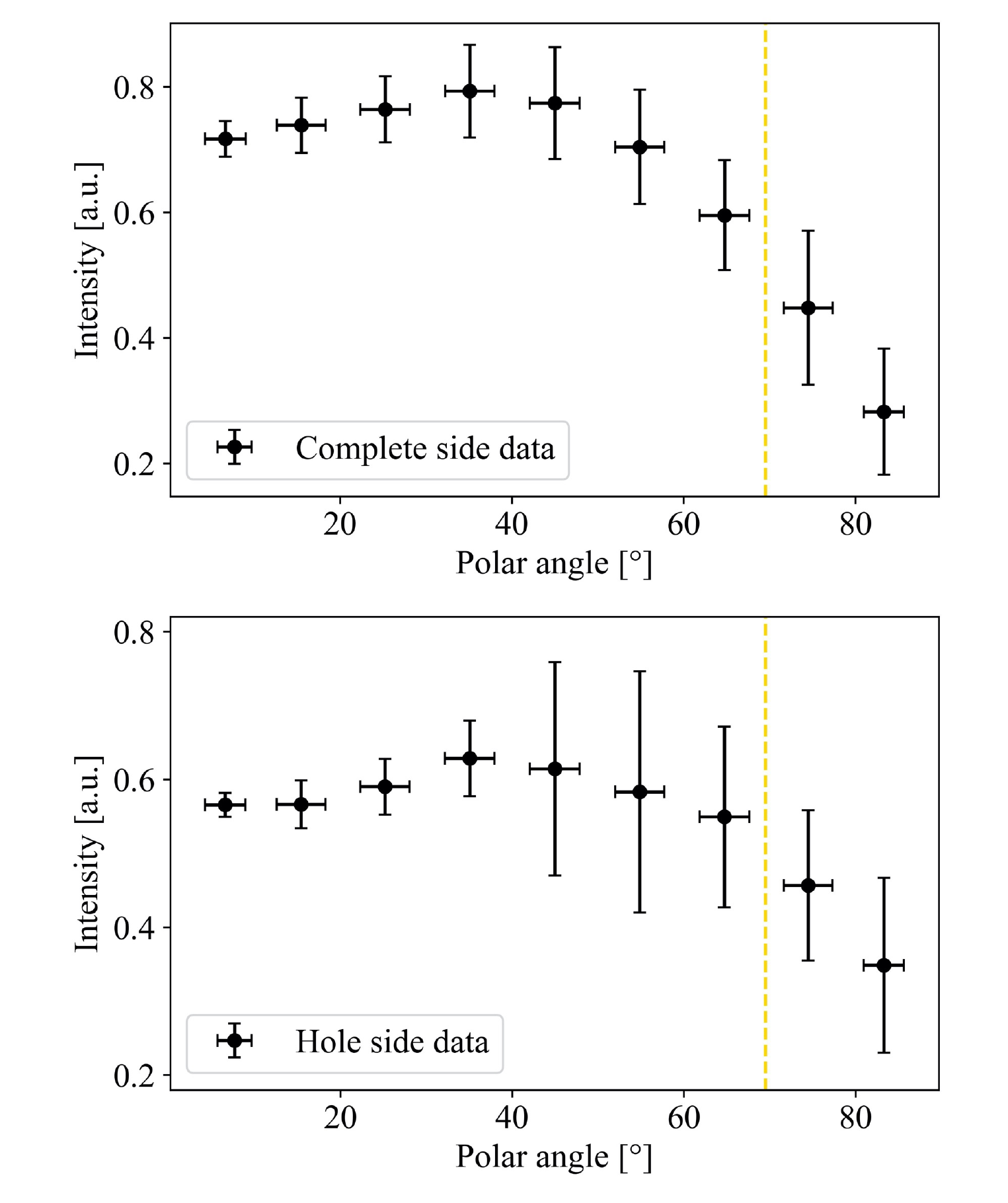}
    \caption{\label{fig_1d_emission_pdf}Light intensity as a function of polar angle. }
\end{subfigure}
\caption{\label{fig_2d_isotropy} 
Emission profile of $460~\mathrm{nm}$ steady light source with a camera exposure time of $0.05~\mathrm{s}$ and a gain of $2$, along with an f/1.8 lens in low-temperature calibration. Figure (a) shows the light-emitting area for the polar angle $\leq 70^\circ$, corresponding to the 3D-printed diffuser coverage. Figure (b) shows the light intensity as a function of the polar angle for the complete hemisphere (top) and hole hemisphere (bottom), derived from (a). The polar angle of $0^\circ$ represents the pole of the light source sphere, while $90^\circ$ indicates horizontal. The standard deviation of gray values and polar angles within each bin is also shown. The gold dash line denotes the polar angle of $70^\circ$.
}
\end{figure}

The emission profile of the $460~\mathrm{nm}$ steady light source is presented in Figure~\ref{fig_2d_emission_pdf}, visualized in Lambert projection. 
The emission area of the light source is shown in color, concentrated within the region of the polar angle $\leq 70^\circ$, which corresponds to the area covered by the 3D printed diffusers. The light emanating from the remaining area, denoted as the white space, is shielded by the 3D printed support used for mounting electronic boards.
The colorbar indicates the normalized gray value. Due to the normalization of the emission profile of two hemispheres, the maximum value of the colorbar is $\sim 0.8$ since there exists a brightness ratio of $1.27$ for the $460~\mathrm{nm}$ steady light source. The two black shadows in the lower hemisphere are attributed to the vacuum port and penetrator on the glass vessel, and the line-shaped shadows in the upper hemisphere are due to cables and fibers encapsulated within the light source.
Figure~\ref{fig_1d_emission_pdf} shows the light intensity of the complete (top) and hole (bottom) hemispheres as a function of the polar angle corresponding to Figure~\ref{fig_2d_emission_pdf}. The polar angles are divided into several bins with a bin width of $\Delta \theta=10^\circ$, and all azimuthal angles of each bin are averaged to visualize the variation of light source brightness with polar angles. The gold dashed line denotes the polar angle of $70^\circ$.

The isotropy of the $460~\mathrm{nm}$ light source is obtained by calculating the average and standard deviation of the gray value of the entire light-emitting area depicted in Figure~\ref{fig_2d_emission_pdf}. 
The ratio of the standard deviation to the mean is used to quantify the anisotropy of the light source.
The resulting anisotropy of the complete hemisphere and the hole hemisphere are $13.6\%$ and $19.1\%$, respectively. The overall anisotropy of the two hemispheres is $19.2\%$.
The anisotropy of the hole hemisphere is mainly caused by the vacuum port and the penetrator.
The calibrated emission profile, which incorporates the anisotropy, is used in the data analysis of the camera system \cite{TianWei:2022}.

\begin{figure}[htb]
    \centering
    \includegraphics[width=0.98\textwidth]{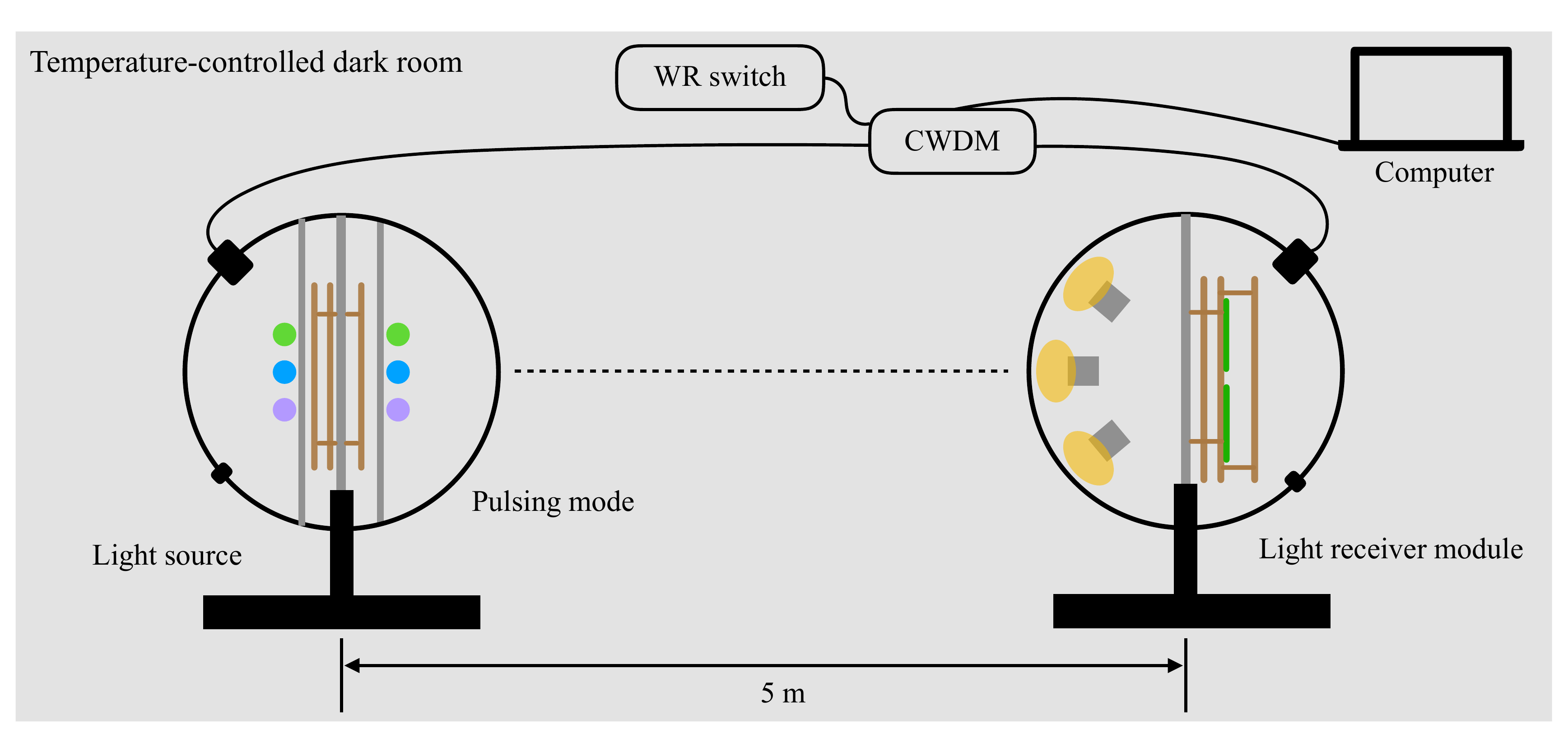}
    \caption{Sketch of setup for low-temperature calibration of the pulsing light source with PMTs in the receiver module. 
    }
    \label{fig_PMT_setup} 
\end{figure}

\subsection{Pulsing light source calibration}
\label{sec:pulsing_light_source_calibation}

We conducted a calibration of the brightness ratio of the pulsing light source using six PMTs in two receiver modules. The laboratory setup used in the calibration is shown in Figure~\ref{fig_PMT_setup}. The light source and the receiver module were positioned opposite to each other in a dark room, with a distance of $5.00\pm 0.03~\mathrm{m}$ between them. The two modules were connected to the computer via a CWDM, and a stand-alone WR switch was also connected to the CWDM to synchronize the clock.

\begin{figure}[htbp]
    \centering
    \includegraphics[width=0.7\textwidth]{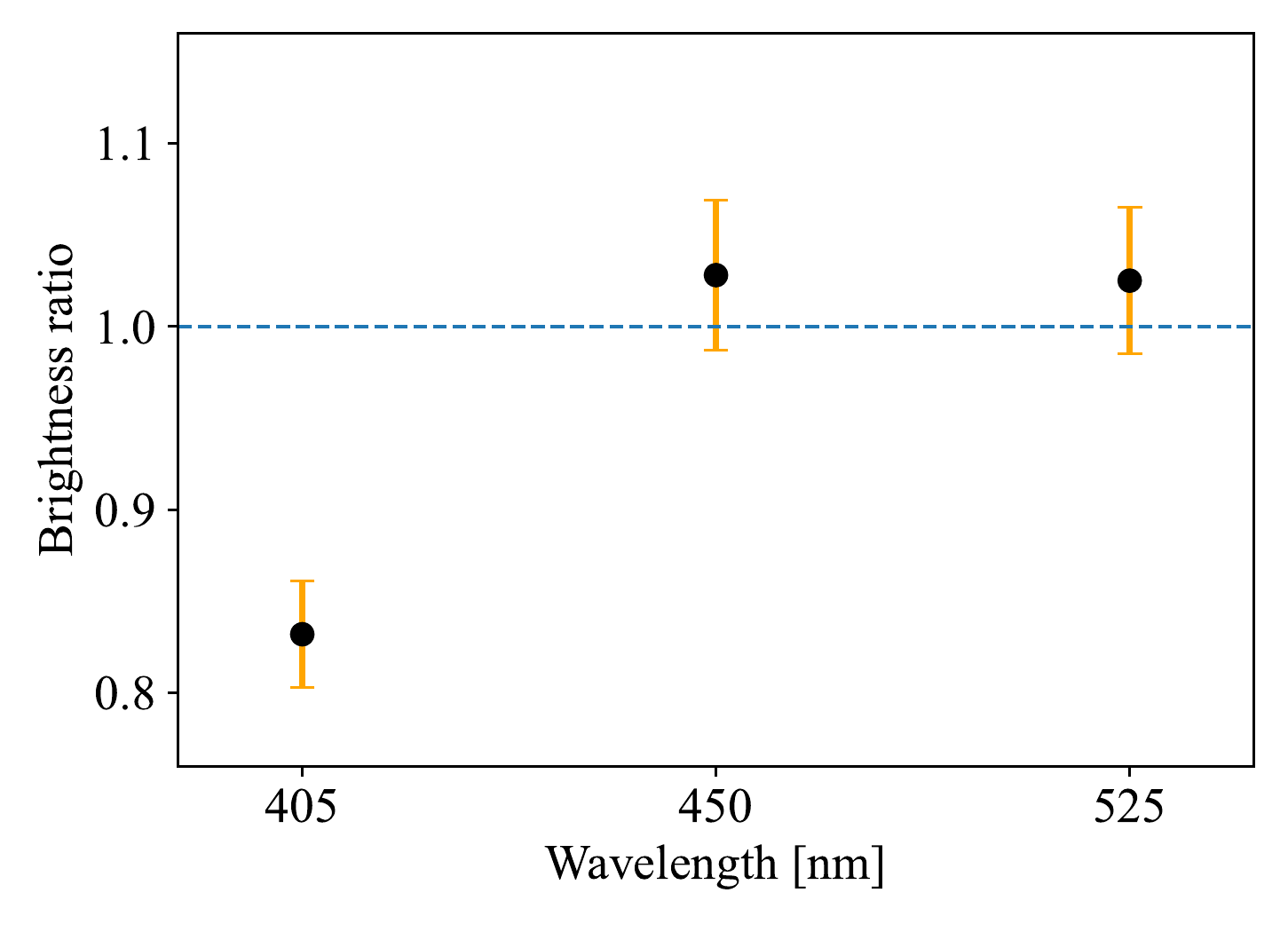}
    \caption{The brightness ratio (complete hemisphere over hole hemisphere) of the pulsing light source as a function of wavelength. The blue dashed line denotes a brightness ratio of 1.}
    \label{fig_pulsing_led_brightness_ratio} 
\end{figure}

The calibrated brightness ratio of the complete hemisphere over the hole hemisphere for each wavelength is illustrated in Figure~\ref{fig_pulsing_led_brightness_ratio}. The brightness ratio of $405~\mathrm{nm}$, $450~\mathrm{nm}$, and $525~\mathrm{nm}$ pulsing light source are $0.832\pm 0.029$, $1.028\pm 0.041$, and $1.025\pm 0.040$, respectively. 
The brightness ratio, as an important input parameter, is used in the analysis of the PMT system. 
While the calibration of the pulsing intensity and emission time profile of the pulsing light source is interrelated with the photon detection efficiency and time response profile of PMTs, this part is discussed in detail in the PMT system paper \cite{Yudi:2022}.

\section{Light source performance in T-REX}
\label{sec: light source performance}

This section presents the performance of the light source during the data collection period of T-REX. The T-REX apparatus was deployed to a depth of $3420~\mathrm{m}$ by an umbilical cable on the research vessel. 
During deployment, the light source emitted $460~\mathrm{nm}$ steady light at depths of $1221~\mathrm{m}$ and $2042~\mathrm{m}$, allowing the camera to capture data for $8$ minutes at each depth.

After deployment, the PMT system collected data first. The pulsing LEDs were powered with a flash frequency of $10~\mathrm{kHz}$. Although we previously calibrated the number of received photoelectrons versus the supply voltage with an attenuator in the lab (see Section~\ref{pulsing_LED_calibration}), it serves as a reference since there could be a disparity between the attenuation of seawater and the attenuation of the single attenuator used in the lab. To obtain reliable SPE signals in the deep-sea experiment, an automatic voltage scan process \cite{Jiannan:2022} was conducted to select an appropriate supply voltage for each pulsing LED. The slow control system scanned the supply voltage of pulsing LEDs ranging from $5~\mathrm{V}$ to $25~\mathrm{V}$ with a step of $1~\mathrm{V}$, collecting $\sim 20,000$ waveforms for each voltage. 
Fast data analysis was conducted to calculate the average number of photoelectrons received by the PMTs at each voltage. The supply voltage was then determined to ensure an average signal expectation of less than $1$ photoelectron per PMT. Specifically, a voltage of $8~\mathrm{V}$ was selected for the $450~\mathrm{nm}$ LED, corresponding to an average of $0.3$ photoelectrons per PMT for each pulse. For the $405~\mathrm{nm}$ and $525~\mathrm{nm}$ LEDs, voltages of $11.5~\mathrm{V}$ and $10~\mathrm{V}$, respectively, were employed, corresponding to an average of $0.6$ photoelectrons per PMT for each pulse.
The entire voltage scanning process lasted $\sim 5~\mathrm{minutes}$. Owing to the automatic voltage scan system, the pulsing light source provided SPE signals to PMTs. 
The PMT system began taking data with the $450~\mathrm{nm}$ light source for $\sim 50~\mathrm{minutes}$, collecting more than $10^7$ photons. Then $10$ minutes of data collection was conducted sequentially for the light source at wavelengths of $405~\mathrm{nm}$ and $525~\mathrm{nm}$.
Throughout the entire PMT data collection period, clock synchronization was achieved using the WR system with sub-nanosecond precision.

After the completion of PMT system data collection, the PMTs were turned off, and the light source was switched from pulsing to steady mode. Five identical steady LEDs of each wavelength were connected in series and powered by a constant current of $20~\mathrm{mA}$. 
The light source emitted steady light with an order of $460~\mathrm{nm}$, $405~\mathrm{nm}$, and $525~\mathrm{nm}$. For each wavelength, the camera captured $\sim 1000$ images of the steady light source.
The entire data collection process of the camera system took $\sim 30~\mathrm{minutes}$. 
Following this, the T-REX apparatus was retrieved.
During the retrieval process, the steady light source continued to emit light at a wavelength of $460~\mathrm{nm}$ for the camera to capture data. Throughout the entire data collection period, the light source performed stably and supplied the expected light to the receiver modules.

\section{Summary}
\label{sec: summary}
This article presents the light source's design, calibration, and performance in the TRIDENT pathfinder experiment.
The light source is designed to emit light in tunable pulsing and steady modes at multiple wavelengths, while a double diffuser structure is used to optimize its isotropy.
The pulsing mode can emit intense, nanosecond-width, and adjustable light pulses, while the steady mode emits isotropic and bright light. A WR system is incorporated in the light source to achieve sub-nanosecond clock synchronization. The pulsing light source provides SPE signals to PMTs with a supply voltage confirmed by an automatic voltage scan process. The isotropy and brightness ratio of the light source are accurately calibrated in the lab to ensure reliable and precise measurements.

The tunable pulsing and steady modes at multiple wavelengths, along with the sub-nanosecond clock synchronization, make this light source design suitable for use in optical experiments that require precise timing and synchronization.
With further optimization, the LED system with two switchable emission modes can become a valuable tool for accurate and rapid calibration, either integrated into the hybrid digital optical module (hDOM) \cite{trident:2022} or stand-alone calibration modules in the future telescope.

\section*{Acknowledgements}


We thank Jun Guo for his valuable comments and insightful discussions to improve this paper.

This work is supported by the Ministry of Science and Technology of China [No. 2022YFA1605500]; Office of Science and Technology, Shanghai Municipal Government [No. 22JC1410100]; and Shanghai Jiao Tong University under the Double First Class startup fund and
the Foresight grants [No. 21X010202013] and [No. 21X010200816].

\bibliography{light_source_2023}

\end{document}